\documentclass[acmsmall,screen]{acmart}

\usepackage{graphicx}%
\usepackage{multirow}%

\usepackage{amsmath,amssymb,amsfonts}%
\usepackage{amsthm}%
\usepackage{mathrsfs}%
\usepackage[title]{appendix}%
\usepackage{xcolor}%
\usepackage{textcomp}%
\usepackage{manyfoot}%
\usepackage{booktabs}%
\usepackage{colortbl}
\usepackage{algorithm}%
\usepackage{algorithmicx}%
\usepackage{algpseudocode}%
\usepackage{listings}%
\usepackage{tcolorbox}
\usepackage{subcaption}
\usepackage{makecell}
\usepackage{array} 
\usepackage{framed}
\usepackage{caption}
\usepackage{subcaption}

\newcommand{\ADRFull}{Architectural Decision Record}
\newcommand{\ADDFull}{Architectural Design Decision}
\newcommand{\Approach}{DRAFT} 
\newcommand{\ApproachFull}{Domain-specific Retrieval Augmented Few-shot Tuning} 
\newcommand{\Decision}{Design Decision}
\newcommand{\Context}{Decision Context}
\newcommand{\finetuning}{fine-tuning}
\newcommand{\finetuned}{fine-tuned}
\newcommand{\Finetuning}{Fine-tuning}
\newcommand{\Finetuned}{Fine-tuned}
\newcommand{\fewshot}{few-shot}
\newcommand{\Fewshot}{Few-shot}
\newcommand{\RAG}{RAFG}
\newcommand{\RAGFull}{Retrieval-Augmented Generation}
\newcommand{\RAFewshotG}{Retrieval Augmented Few-shot Generation}

\definecolor{highlight}{RGB}{220,250,220}
\definecolor{border}{RGB}{55,135,55}
\definecolor{best}{RGB}{198,239,206}   
\definecolor{worst}{RGB}{255,199,206}  

\newcommand{\rebuttal}[1]{\textcolor{black}{#1}}

\newenvironment{rebuttalblue}{
    \color{black}
}

\AtBeginDocument{%
  }

\begin{document}

\title{\textit{\Approach}-ing Architectural Design Decisions using LLMs}


\author{Rudra Dhar}
\affiliation{%
  \institution{IIIT Hyderabad}
  \city{Hyderabad}
  \country{India}}
\email{rudra.dhar@research.iiit.ac.in}

\author{Adyansh Kakran}
\affiliation{%
  \institution{IIIT Hyderabad}
  \city{Hyderabad}
  \country{India}}
\email{adyansh.kakran@research.iiit.ac.in}

\author{Amey Karan}
\affiliation{%
  \institution{IIIT Hyderabad}
  \city{Hyderabad}
  \country{India}}
\email{amey.karan@research.iiit.ac.in}

\author{Vasudeva Varma}
\affiliation{%
  \institution{IIIT Hyderabad}
  \city{Hyderabad}
  \country{India}}
\email{vv@iiit.ac.in}

\author{Karthik Vaidhyanathan}
\affiliation{%
  \institution{IIIT Hyderabad}
  \city{Hyderabad}
  \country{India}}
\email{karthik.vaidhyanathan@iiit.ac.in}

\renewcommand{\shortauthors}{Dhar et al.}

\begin{abstract}
Architectural Knowledge Management (AKM) is crucial for software development  and \ADRFull s (ADR), an important artifact of AKM provide a structured approach to capture \ADDFull s (ADD). However, their adoption is limited due to the manual effort involved and insufficient tool support. Prior work shows that LLMs can support ADD generation, though basic prompting yields suboptimal results.

This paper explores how LLMs can support architects in two complementary tasks: generating ADDs from Decision Contexts, and generating complete ADRs from titles. We propose \ApproachFull~(\Approach), a novel approach that integrates \RAFewshotG~with \Finetuning~to generate. \Approach~operates in two phases: an offline phase that fine-tunes a model using retrieval-augmented prompts derived from historical ADRs, and an online phase that generates decisions or complete ADRs by combining retrieved examples with the \finetuned~model.

Evaluated on 4344 ADRs across multiple LLMs using standard NLP metrics, \Approach~consistently produces outputs better aligned with real world ADRs. \rebuttal{A human evaluation with the authors and external architects, however, shows no uniformly preferred approach.} \Approach~also maintains practical efficiency, reduces dependence on proprietary models, and addresses organizational concerns around privacy, resource constraints, and inference costs.
\end{abstract}




\maketitle

\section{Introduction}\label{sec:introduction}

\textit{Architectural Knowledge Management (AKM)} refers to the systematic capture, storage, and reuse of \textit{Architectural Knowledge (AK)} within software projects or an organization. This knowledge typically includes architectural styles, design patterns, quality attributes, and critical \Decision. AKM addresses the key challenge of architectural knowledge vaporization, the gradual loss of valuable architectural knowledge over time as defined by Tofan et al. \cite{10.1007/978-3-642-39031-9_17}. Effective AKM ensures traceability, supports collaboration and reuse, and improves communication, learning, and documentation, ultimately contributing to project success.

Despite its recognized importance, AKM has long suffered from limited adoption. As studied by Tang et al. \cite{TANG2010352} various tools have been developed to support AKM processes, but these tools have not been sufficient.
As noted by Capilla et al. \cite{CAPILLA2016191}, current efforts fall short in effectively capturing and documenting AK. 
\rebuttal{While recent efforts have developed semantic modeling tools and version controlled repositories to support AKM processes and enable AI based analysis \cite{Karetnikov}, the manual overhead of documenting decisions remains a significant barrier.}
This gap highlights the need for more research into automating knowledge capture to ease the burden on architects and development teams.

Jansen et al. \cite{ADD} characterize software architecture as a set of key \textbf{\ADDFull s (ADDs)} that shape a system’s structure and evolution. Architects must analyze complex contexts, evaluate alternatives, and weigh trade-offs to reach sound decisions, which is a cognitively demanding task.

As software systems evolve and development teams change, the need to preserve architectural decisions and their underlying rationale becomes increasingly apparent. Decisions that are not explicitly captured are difficult to revisit, justify, or adapt, which can negatively impact maintainability and long-term system evolution. To address this challenge, Architectural Decision Records (ADRs) were introduced as a lightweight documentation mechanism to record architectural decisions together with their context, rationale, alternatives, and consequences.

Ahmeti et al. \cite{10.1007/978-3-031-70797-1_22} confirm the clear benefits of using ADRs. However, their adoption has been low in practice. As studied by Buchgeher et al. \cite{adrs_data} this is largely due to the high manual effort required to document decisions, the lack of adequate tool support, interruptions to the design process caused by documentation overhead and uncertainty about which aspects of AK should be documented.
While \Decision~gain clarity through documentation, documenting them imposes additional effort at a time when architects are already cognitively engaged in design reasoning. As a result, the combined burden of reasoning about decisions and expressing them in a structured, reusable form becomes a key obstacle to sustained ADR usage.

This motivates viewing architectural decision support as two complementary, interdependent tasks. The first assists architects during decision exploration by supporting the articulation of candidate decisions given a context. The second supports the transformation of sufficiently clarified decisions into well structured ADRs.

Recent advances in \textbf{Large Language Models (LLMs)} have opened up new possibilities for automated documentation, including the generation of architectural knowledge artifacts by Hou et al. \cite{10.1145/3695988}. While LLMs show potential in language understanding and documentation, challenges remain in software engineering tasks, including privacy, computational requirements, token consumption costs, output quality issues, and unreliability for complex architectural reasoning tasks, as observed in multiple studies \cite{10.1145/3639476.3639764,10449667,YAO2024100211, 10.1145/3786152.3788588} .

Specific research on the use of LLMs for ADR generation is still limited. To this end, we conducted an exploratory empirical study of whether LLMs can effectively generate ADRs \cite{dhar2024llmsgeneratearchitecturaldesign}. While the goal of generating entire ADRs from a codebase remains a future work, the focus of this study was on utilizing LLMs to generate \Decision s from \Context s as these are recognized as the core components of any ADR \footnote{\url{https://docs.aws.amazon.com/prescriptive-guidance/latest/architectural-decision-records/adr-process.html}}.

Our study showed that LLMs can generate reasonable \Decision s, but the outputs did not consistently match the quality of human architects.
We also observed that while the performance of certain LLMs improved in a \textbf{\fewshot} setting, the overall phenomenon lacked generalization and remained inconclusive. Since the \fewshot~samples were the same for every input, they were not very helpful in cases where the examples were unrelated to the input due to the huge variety in \Decision s.
Moreover, \finetuned~LLMs exhibited improved capability in generating \Decision s. We concluded that compact fine-tuned LLMs, which require minimal infrastructure for hosting, had potential to be used as substitute for extensive and proprietary LLMs in scenarios where privacy and hardware infrastructure, and cost is a concern.

Building on these insights, this work supports architects across broader lifecycle of architectural decision-making, from formulation to documentation. We focus on two complementary tasks that reduce both cognitive and documentation effort. 

\noindent \textbf{Task 1 - Decision Making Support (Context → Decision):} Given a Decision Context describing an architectural problem or situation, generate appropriate Design Decisions to help architects evaluate options and make informed choices. This task assists architects during the decision-making process by providing AI-generated decision alternatives that can serve as starting points for evaluation and discussion.

\noindent \textbf{Task 2 - Complete ADR Generation (Title → Body):} Given an ADRs Title (typically written after a decision has been made), generate a complete ADR that properly documents the context, rationale, consequences, alternatives considered, and other relevant information. This task is instantiated as (Title → Body) where \textbf{Body} refers to the complete ADR except the Title. This task assists architects in creating comprehensive documentation without the time-intensive manual effort traditionally required.

To address these challenges, we explored \RAGFull~(RAG) introduced by Lewis et al. \cite{lewis2021retrievalaugmentedgenerationknowledgeintensivenlp}, which combines retrieval of relevant information with generative AI, to generate more accurate and relevant responses.
In particular, we used \RAFewshotG~(\RAG), introduced by Izacard et al. \cite{izacard2022atlasfewshotlearningretrieval}, where the output is generated using a \fewshot~prompt where the examples are retrieved from a database.
Our experiments showed that \RAG~does improve the ability of LLMs to generate \Decision.

Inspired by our previous work \cite{dhar2024llmsgeneratearchitecturaldesign} and observations, we came up with a novel approach \rebuttal{combining the existing NLP techniques} called \textbf{\ApproachFull~(\Approach)}. The approach uses the concept of \RAFewshotG, along with \finetuning.
It broadly has 2 phases - offline and online.

In the offline phase, a foundational model is \finetuned~to produce outputs from given inputs and a few similar input-output pairs. These similar input-output pairs are retrieved from a vector database. In the online phase, users provide input, which is used to retrieve similar input-output pairs and generate the target output using the fine-tuned LLM. \Approach~is designed to work for both tasks, adapting to the specific requirements of decision generation, and documentation.

\rebuttal{It is important to note that DRAFT is not intended to autonomously replace the architect's decision-making process. Recent software engineering studies highlight that fully autonomous LLMs remain unreliable for complex architectural reasoning \cite{10.1145/3786152.3788588}. Instead, DRAFT acts as a drafting assistant or copilot, keeping the human in the loop to reduce cognitive overhead during alternative exploration and mitigate documentation friction.}

\rebuttal{Through extensive evaluation on a dataset of 4,344 ADRs, we demonstrate that \Approach~aligns more closely with real-world ADRs than existing approaches on automated metrics while maintaining efficiency, although a complementary human evaluation shows that practitioners may value richer rationale over this alignment.
}
Our findings suggest that \Approach~offers a practical solution to assist architects in drafting ADRs, especially for organizations facing privacy and infrastructure constraints.
The source code for all the experiments alongside the data used is available on the replication package \footnote{\url{https://zenodo.org/records/23438622}}.

The rest of this paper is organized as follows. Section \ref{sec:background} provides background information on ADRs, and text generation techniques. Section  \ref{sec:Overview} highlights the motivation and presents an overview of \Approach, followed by a detailed explanation in Section \ref{sec:Approach}. Section \ref{sec:results} evaluates the performance of \Approach~in comparison to existing approaches. Section \ref{sec:Discussion} discusses key lessons learned and their broader implications. Section \ref{sec:threats_to_validity} outlines potential threats to validity, while Section \ref{sec:related_work} reviews related work. Finally, Section \ref{sec:conclusion} concludes the paper.

\section{Background and Motivation}\label{sec:background}

\subsection{\ADRFull~(ADR)}\label{sec:ADR}




   
    
    




\begin{figure}[ht]
\centering
\begin{tcolorbox}[colback={highlight}, colframe={border}]
\begin{rebuttalblue}
External Facing API Design and Implementation\\
    {\scriptsize
    \par
Date: 2021-09-23

\textbf{Status}

Accepted

\textbf{Context}

The Office of Head Start is expanding data sharing across internal systems. To support this, TTA Hub would like to make documented APIs available to external systems in reliable, consistent, and secure ways.

\textbf{Options Considered}

To achieve this, two paths were considered.

\textbf{Expose existing internal APIs}

TTA Hub has a robust set of APIs already implemented to support the React frontend.

Pros:

\begin{itemize}
\item already implemented
\item well documented
\end{itemize}

Cons:

\begin{itemize}
\item API is highly coupled to TTA Hub UI needs, will become more difficult to change rapidly
\item authentication is based on sessions stored in browser cookies
\item authorization fully limited to existing TTA Hub roles
\end{itemize}

\textbf{Create a new API layer for external partners}

Creating a new API layer would de-couple the data needs from the TTA Hub UI needs.

Pros:

\begin{itemize}
\item existing API can continue to be optimized for UI considerations
\item new API can be more rigidly backwards-compatible
\item authentication schemes can be easily customized for each use
\item authorization can be scope-based and integrated into the full HSES OAuth implementation
\end{itemize}

Cons:

\begin{itemize}
\item two API endpoints will need to be maintained through data model changes
\end{itemize}

\textbf{Decision}

A new API layer has been chosen as the better set of trade offs.

To distinguish between the API flavors, the external API will be namespaced under the /api/v1 route path.

\textbf{Authentication}\\
The initial API client also utilizes HSES as a single-sign-on solution, so API authentication will be done by passing an HSES token via the Authentication HTTP header. TTA Hub will then validate that token with HSES to verify the user.

\textbf{JSON Format}\\
To promote consistency, the external facing API will conform to the JSON:API schema as much as possible. The first endpoints will include objects in attributes that may eventually migrate to relationships. This trade-off has been made due to the exponential growth of required API endpoints when each object is represented in the relationships section.

\textbf{Consequences}

The result of this decision will be a stable external API with simple authentication logic, while maintaining agility in the UI-driven internal API. The risk comes from undertaking and maintaining a second representation of the TTA Hub data.
\par
    }
    \end{rebuttalblue}
\end{tcolorbox}
\caption{Sample ADR illustrating the standard Status, Context, Decision, and Consequences sections.}
\label{fig:ADR}
\end{figure}

Software architecture is fundamentally a collection of key \textit{\Decision s} as defined by \cite{ADD}. \rebuttal{Elevating these decisions to first class status through explicit documentation makes the architecture development process more structured and transparent. \cite{DemystifyingArchitecture}}
An \textit{\ADRFull~(ADR)} is a lightweight document used in software development to capture key architectural decisions made throughout a project's lifecycle.
\rebuttal{ADRs originated as a practical industry standard, first introduced in a 2011 blog post by practitioner Michael Nygard\footnote{\url{https://www.cognitect.com/blog/2011/11/15/documenting-architecture-decisions}}, rather than emerging from a foundational academic research paper. However, the concept has since been formalized and studied academically \cite{Zimmermann}. }
This document includes details about the \textit{context} of the problem, the \textit{decision} reached, the expected outcomes of the decision, any pertinent references, and the status of the decision.
\rebuttal{Modern ADRs use structured Markdown templates that integrate with version control, bridging design and implementation \cite{madr}.}
ADRs enhance transparency, encourage collaboration, and maintain the historical context of architectural decisions, ensuring that decision-making is well-informed. A sample ADR is shown in Figure  \ref{fig:ADR}.

ADRs are tightly coupled with architectural decision making. Architects iteratively explore alternatives, articulate decisions, and refine their understanding of problem and solution. Documenting decisions in ADRs externalizes reasoning, preserves knowledge, and often triggers further reflection and improvement.
From an AKM perspective, ADRs serve dual roles: they capture architectural decisions and support ongoing reasoning. While documentation adds effort, it is essential for stabilizing and sharing knowledge.
The key challenge is to support both decision making and documentation with minimal friction and cognitive overhead. We therefore distinguish two complementary forms of support aligned with AKM workflows:
\begin{itemize}
    \item Assistance during exploration and formulation of candidate decisions in context,
    \item Help with expressing finalized decisions in structured ADR format.
\end{itemize}
These are not strictly sequential, rather they reflect different emphases within an iterative process where decisions and their documentation co-evolve.

\medskip
\noindent To support both stages, we define two complementary tasks:
\begin{itemize}
    \item \textbf{Task 1 -- Decision Making Support (Context $\rightarrow$ Decision):} \\
    Generating \Decision~from Decision Context of an ADR.\\
    \emph{Input:} Decision Context (C) \\
    \emph{Output:} Design Decision (D)

    \item \textbf{Task 2 -- Complete ADR Generation (Title $\rightarrow$ Body):} \\
    Generating complete ADRs (instantiated as Generating ADR bodies from it's Title.)\\
    \emph{Input:} Decision Title (T) \\
    \emph{Output:} ADR Body (B)
\end{itemize}

\textit{Throughout the remainder of this paper, we use} \textbf{Task~1 (Context $\rightarrow$ Decision)} \textit{as the
primary illustrative example when explaining concepts, algorithms, and figures.
The same principles, techniques, and processes apply identically to}
\textbf{Task~2 (Title $\rightarrow$ Body)}, \textit{with the only difference being the substitution
of input--output pairs.}

\medskip

\noindent In this paper:
\begin{itemize}
    \item Decision Context is referred to as \textbf{Context} and denoted by $C$
    \item Design Decision is referred to as \textbf{Decision} and denoted by $D$
    \item Title is denoted by $T$
    \item Body is denoted by $B$
\end{itemize}

\begin{rebuttalblue}
\subsection{Prompting}



Prompting elicits outputs by providing textual input that guides the model toward a desired response. In \textbf{zero-shot} prompting, the model relies on its pre-trained knowledge to infer and generate a decision based solely on the given context.

For \textbf{Task 1}, this can be represented as:
$ D \leftarrow LLM(C) $
where $C$ is the \Context~provided as input and $D$ is the \Decision~generated as output.
For \textbf{Task 2}, the formulation would be:
$ B \leftarrow LLM(T) $
where T is the Title provided as input and B is the ADR Body generated as output.

\textbf{\Fewshot}~\cite{brown2020languagemodelsfewshotlearners} prompting introduces a limited set of sample pairs within the prompt to provide the LLM with implicit task-specific knowledge. By presenting these exemplars, the model is guided toward recognizing patterns and generating outputs that align with prior examples.

For Task 1 
(Context → Decision), this can be formulated as:
\[ D \leftarrow LLM(\{(C_1, D_1), (C_2, D_2), \dots, (C_k, D_k), C\}) \]
where 
$k$ denotes the number of exemplars included in the prompt, and $(C_i, D_i)$ represents an example pair of Context and Decision, with $i$ going from 1 to $k$. The presence of these exemplars facilitates more accurate and contextually relevant outputs compared to zero-shot prompting.

For \textbf{Task 2 (Title → Body)}, the formulation would be:
\[ B \leftarrow LLM(\{(T_1, B_1), (T_2, B_2), \dots, (T_k, B_k), T\}) \]
where $(T_i, B_i)$ represents an example pair of Title and Body.



\subsection{\RAFewshotG} \label{sec:rag}

\RAGFull~(RAG) introduced by Lewis et al. \cite{lewis2021retrievalaugmentedgenerationknowledgeintensivenlp} combines LLMs with external retrieval to improve contextual accuracy and factual reliability. Multiple studies have shown complex Retrieval architectures such as Knowledge Graphs \cite{barron2024domainspecificretrievalaugmentedgenerationusing}, Re-ranking \cite{hui2022retrievalaugmentationt5reranker}, etc. improving the performance of RAG models. A comprehensive survey of RAG techniques is provided by Gao et al. \cite{gao2024retrievalaugmentedgenerationlargelanguage}. 
In this study we use \textbf{\RAFewshotG}, which is a combination of \fewshot~prompting and RAG \cite{izacard2022atlasfewshotlearningretrieval}. Rather than relying on a static, predefined set of \fewshot~examples, this approach retrieves contextually similar examples from a knowledge base, such as a vector database (VDB). This method improves the model’s ability to generate context-aware and semantically relevant responses.
Here is a breakdown of how it works:






\textbf{Embedding and Storage:} Input textual contexts are converted into high-dimensional vector embeddings,
$ v_C \leftarrow f_{\text{embed}}(C) \in \mathbb{R}^d $
and stored in a Vector Database (VDB) such as FAISS (Facebook AI Similarity Search) \cite{douze2025faisslibrary} alongside their corresponding decisions.


\textbf{Retrieval:} For a new input context $C$, a similarity search retrieves the top-$k$ most relevant historical pairs:
\[
f_{\text{search}}(v_{C_q}, VDB) \rightarrow \{ (C_1, D_1), (C_2, D_2), \dots, (C_k, D_k) \}
\]


\textbf{Augmented Generation:} The retrieved context-decision pairs are combined with the input \Context~to construct a \fewshot~prompt:
\[
P = \{ (C_1, D_1), (C_2, D_2), \dots, (C_k, D_k), C \}
\]
This prompt is then passed to the LLM to generate the final \Decision~$D$
\[
D_q \leftarrow LLM(P)
\]
\end{rebuttalblue}

\begin{figure*}[ht]
    \centering
    \includegraphics[width=0.7\textwidth]{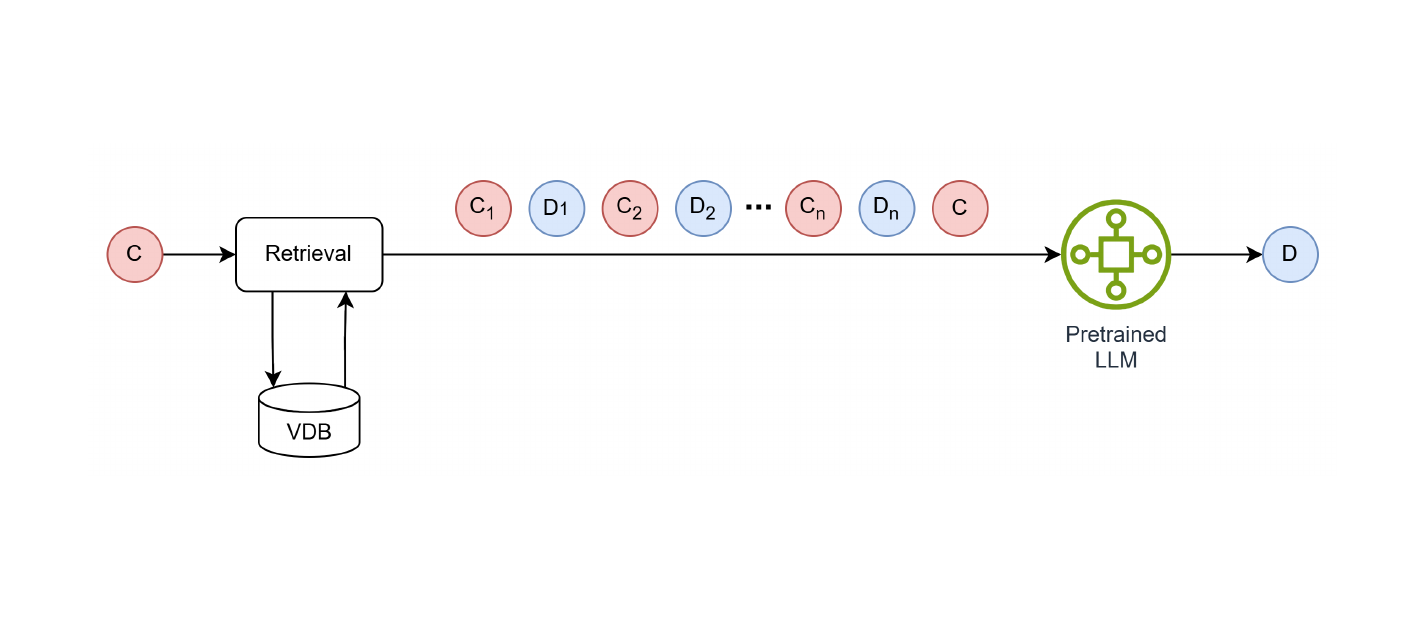}
    \caption{\RAFewshotG~(\RAG) pipeline at runtime.}
    \label{fig:layerRAG}
\end{figure*}

Figure \ref{fig:layerRAG} illustrates the runtime process of \RAG, showing the sequence of events that occur when a \Context~is received and processed to produce a \Decision. Here is a step-by-step breakdown:

\begin{enumerate}
    \item The input context $C$ is embedded as a vector $v_C$, which is a query-appropriate representation.
    \item This representation is used to retrieve similar context-decision pairs from the VDB. \[(C_1,D_1), (C_2,D_2), \ldots, (C_k, D_k)\]  
    \item A \fewshot~prompt is created with the retrieved $C-D$ pairs and the original context $C$.
    \item Taking this prompt as input, the LLM outputs the decision $D$.
\end{enumerate}

For \textbf{Task 2 (Title → Body)} generation, the process is identical. 
The VDB contains Title-Body pairs $\{(T_i, B_i)\}$, and given an input Title $T$, the 
system retrieves similar $\{ (T_1, B_1), (T_2, B_2), \dots, (T_k, B_k) \}$ pairs and constructs 
a few-shot prompt to generate the Body $B$.

\textbf{Limitation:} Although \RAG~enhances contextual accuracy, it has inherent limitations. While the retrieval mechanism improves factual grounding, the model itself does not internalize domain-specific knowledge beyond what is present in the retrieved examples. Hence, other techniques like \finetuning, are required to optimize its performance in domain-specific applications like generating ADDs.

Please note that in this paper the term `model' or `LLM' specifically refers to generative LLMs, unless `embedding model' is explicitly mentioned.


\subsection{\Finetuning}\label{sec:Finetuning}
\Finetuning~is a pivotal technique in machine learning, particularly within the domain of \textbf{transfer learning}~\cite{howard-ruder-2018-universal} . It involves adapting pre-trained foundational models to new tasks or datasets by refining their parameters with domain-specific data \cite{vaswani2023attentionneed, howard-ruder-2018-universal}. Unlike prompting and \RAG, which rely on externally provided context at inference time, \finetuning~embeds domain knowledge directly into the model’s parameters, improving its ability to generate domain-specific responses.
\[ \theta_{\text{fine}} = \theta_{\text{pre}} + \Delta \theta \]
where $\theta_{pre}$ represents the \textbf{pre-trained} model and $\Delta \theta$ represents the adaptations necessary for the new task.

\Finetuning~is widely applied across NLP. For example, pre-trained LLMs, like BERT or GPT, are fine-tuned for tasks like sentiment analysis or question answering.

\textbf{Parameter-Efficient Fine-Tuning (PEFT)} methodologies address the computational burden of adapting large pre-trained models by selectively updating a minimal subset of model parameters \cite{xu2023parameterefficientfinetuningmethodspretrained, han2024parameterefficientfinetuninglargemodels}.
This contrasts with \textbf{full parameter \finetuning}, which necessitates the gradient-based optimization of all model weights, thereby incurring substantial computational and memory overhead.  PEFT techniques, such as adapter modules or \textbf{Low-Rank Adaptation (LoRA)} \cite{hu2022lora}, which introduces low-rank matrices to approximate weight updates, achieve comparable performance to full parameter \finetuning~while significantly reducing the number of trainable parameters.

\begin{figure}[ht]
    \centering
    \includegraphics[width=0.3\textwidth]{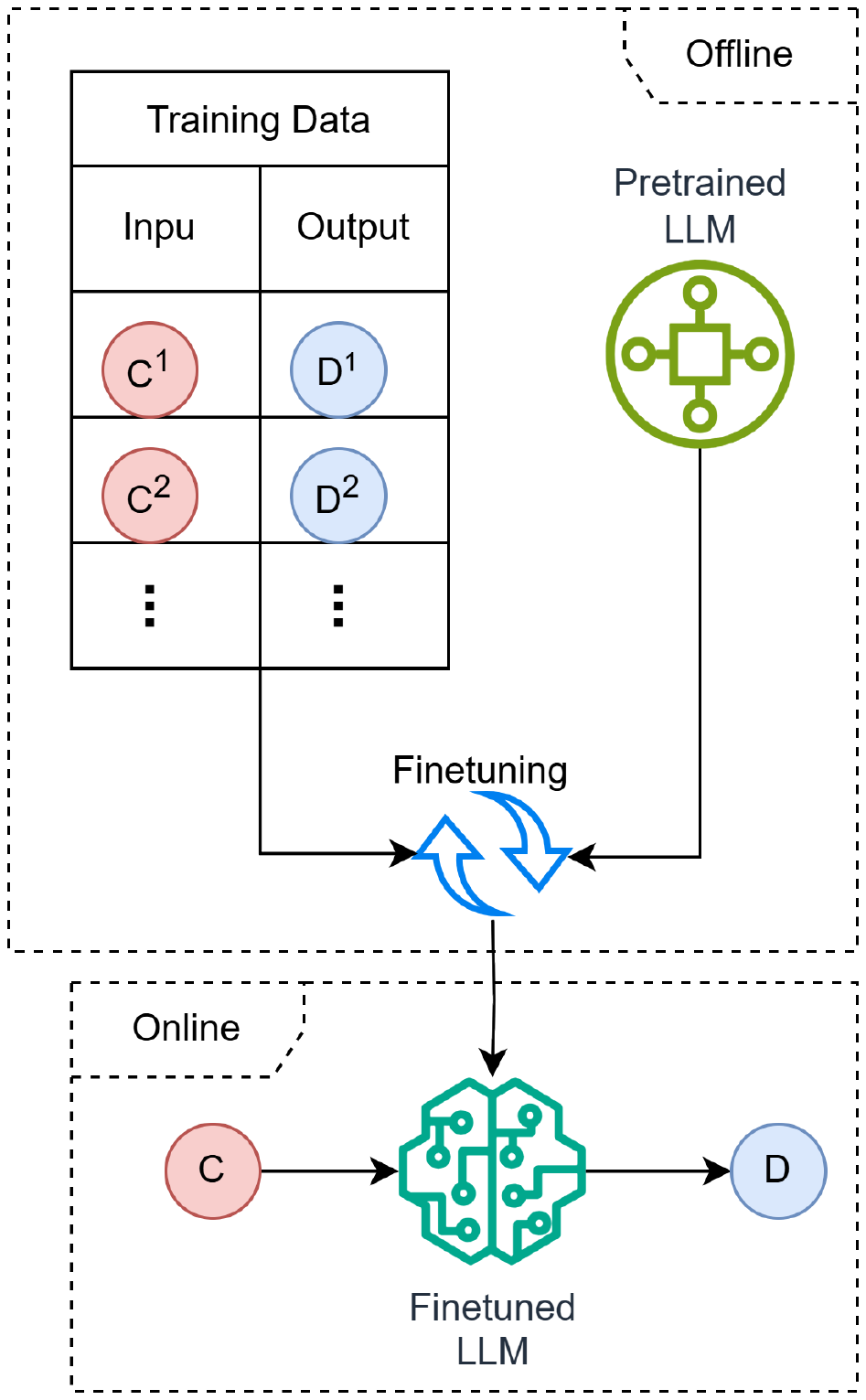}
    \caption{\Finetuning~pipeline showing the offline training phase on context–decision pairs and the online inference phase using the \finetuned~model.}
    \label{fig:layerFinetune}
\end{figure}

Figure \ref{fig:layerFinetune} depicts how \finetuning~can be used to generate \Decision~from a given \Context~(Task 1), which involves two distinct operational phases:

\vspace{5cm}
\textbf{Offline Phase:}\vspace{-4pt}
\begin{enumerate}
    \item \textbf{Dataset Preparation:} Collection and preparation of a dataset containing context-decision pairs $(C^1,D^1), (C^2,D^2), ..., (C^n, D^n)$.
    \item \textbf{Model Training:} The Foundation LLM is \finetuned~on the dataset where $C^i$ is provided as input and $D^i$ is the expected output.
\end{enumerate}

\textbf{Online Phase:}
During inference, the \finetuned~model processes a new architectural context $C$ and generates a decision $D$ using the adapted parameters.

For Task 2 (Title $\rightarrow$ Body), the fine-tuning process is identical. The dataset contains Title-Body pairs $(T^1,B^1), (T^2,B^2), ..., (T^n, B^n)$, and the model is fine-tuned to generate Body $B$ from input Title $T$.

\textbf{Limitation:} While \finetuning~enhances performance, it has inherent constraints. Unlike \RAG, which dynamically fetches relevant examples from external sources, \finetuning~is limited to the knowledge encapsulated during the training process and is constrained by the scope of the training data, making it less adaptable to evolving information.

Please note \Finetuning~is often referred as training in this paper as \finetuning~is a type of training.

\section{\Approach~- Overview}\label{sec:Overview}

In the paper we introduce \Approach, a novel approach that combines \RAG~\fewshot~and \Finetuning~to overcome their individual limitations as mentioned in the previous section. Integrating retrieval with task-specific optimization enables more robust ADD generation by providing both access to external knowledge and targeted model adaptation.


\begin{rebuttalblue}
The development of \Approach~was motivated by two observations. First, LLMs perform better when few-shot examples are provided in their context window~\cite{brown2020languagemodelsfewshotlearners}, but this ability emerged mainly in very large models (more than 100 billion parameters~\cite{brown2020languagemodelsfewshotlearners}) that a small organization cannot host. Second, prompting and RAFG only condition a frozen model on its input and cannot change how the model writes; without adaptation, prompted models produce decisions up to 9$\times$ longer than human-written ones (Section~5.7). Fine-tuning instead updates the model parameters, so that the model internalizes the structure, brevity, and vocabulary of real ADRs, and our previous study showed that it improves ADD generation~\cite{dhar2024llmsgeneratearchitecturaldesign}. Hence, fine-tuned smaller LLMs, which can be deployed on organizational servers, may yield ADDs comparable to those of larger LLMs.
\end{rebuttalblue}

This insight led us to design a novel approach that trains the LLM to perform \fewshot~learning (from sample ADRs) more effectively within a domain (of ADRs). This approach was hypothesized to enable smaller-scale LLMs to generate high-quality ADDs while addressing data privacy concerns and computational constraints faced by smaller organizations. Moreover, \Approach~supports continuous improvement by allowing updates to its VDB.

DRAFT is designed to support both tasks described in Section \ref{sec:ADR}:
\begin{itemize}
    \item \textbf{Task~1 (Context $\rightarrow$ Decision)}: Generating \Decision~from Decision Contexts.
    \item \textbf{Task~2 (Title $\rightarrow$ Body)}: Generating complete ADRs from it's Titles.
\end{itemize}

Both tasks follow the same underlying architecture and process. For clarity, we explain DRAFT using \textbf{Task~1 (Context $\rightarrow$ Decision)} as the running example.
\rebuttal{However the approach applies similarly to Task~2. The main difference is the input-output pairs: instead of context-decision pairs $(C, D)$, Task~2 uses title-body pairs $(T, B)$. Additionaly for prompt engineering, Task 1 employs system instructions demanding a concise decision sentence, while Task 2 leverages expanded instructions detailing required Markdown sub-headers.}

\begin{figure*}[ht]
    \centering
    \includegraphics[width=1\textwidth]{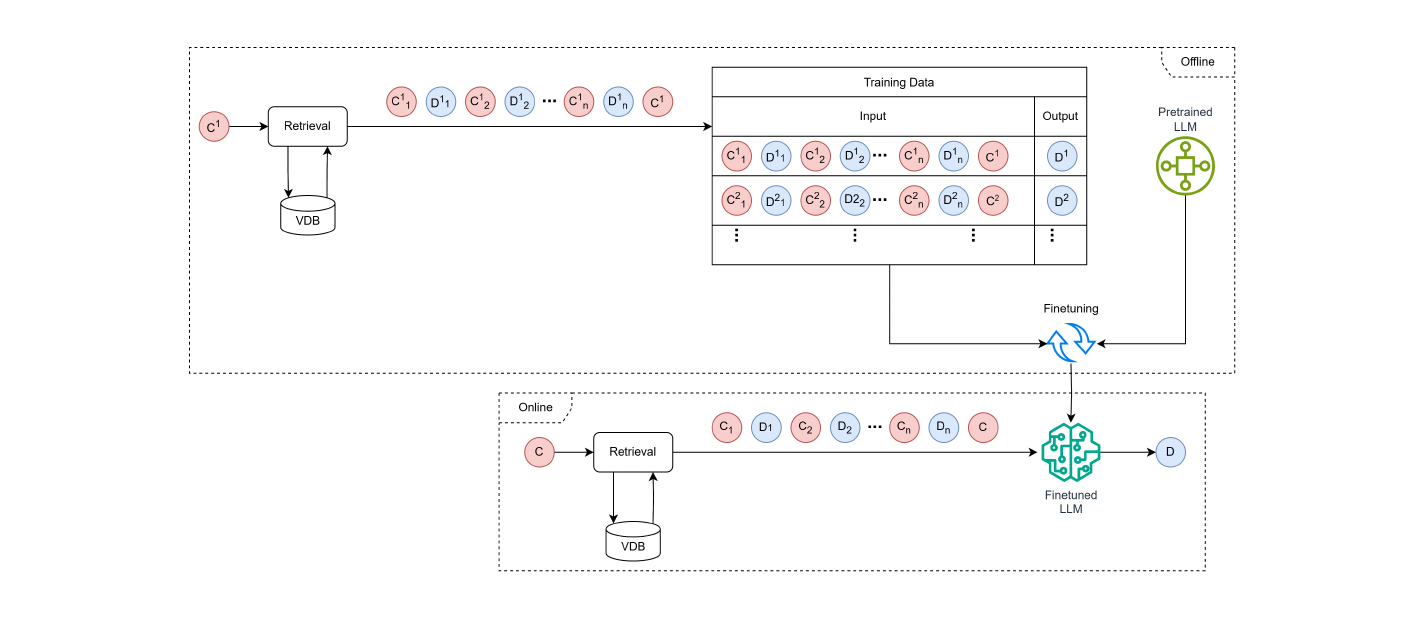}
    \caption{overview of \Approach, illustrating both the offline component (retrieval augmented dataset construction and \finetuning) and the online component (retrieval-augmented inference using the \finetuned model)}
    \label{fig:layerDRAFT}
\end{figure*}

As demonstrated in Figure \ref{fig:layerDRAFT}, \Approach~has an offline and an online component.

\vspace{1pt}
\noindent \textbf{Offline Component:}
\begin{enumerate}

    \item \textbf{Dataset Creation:}
    \begin{enumerate}
        \item The input context $C^i$ is embedded to a vector $v_{C^i}$ which is a query-appropriate representation.
        \item This representation is used to search and retrieve similar context-decision pairs from the VDB.
        \[ (C^i_1, D^i_1), (C^i_2, D^i_2), \dots, (C^i_n, D^i_n) \] 
        \item A \fewshot~prompt is created with the retrieved $C-D$ pairs and the original context $C^i$.
        \[
        P^i = \{(C^i_1, D^i_1), (C^i_2, D^i_2), \dots, (C^i_n, D^i_n),C^i\}
        \]
        \item Creation of training instances that contain this prompt $P^i$ and decision $D^i$ : $\{P^i, D^i\}$
    \end{enumerate}

    \item \textbf{\Finetuning:} The foundation LLM is \finetuned~on the dataset where $P^i$ is provided as input and $D^i$ is the expected output.

\end{enumerate}

\noindent \textbf{Online Component:}
\begin{enumerate}
    \item Given a Context $C^i$, similar ADRs (Context-Decision pairs) are retrieved. 
    \[ (C^i_1, D^i_1), (C^i_2, D^i_2), \dots, (C^i_n, D^i_n) \]
    \item Creation of prompt that contains the context $C^i$ and retrieved context-decision pairs:
    \[
    P^i = \{(C^i_1, D^i_1), (C^i_2, D^i_2), \dots, (C^i_n, D^i_n),C^i\}
    \]
    \item The fine-tuned model receives the prompt $P$ and generates decision $D$ using the adapted parameters.
\end{enumerate}

For Task 2 (Title → Body), the algorithms, system architecture, and all processing steps remain the same. Simply replace:
\begin{itemize}
    \item \emph{Context} $(C)$ with \emph{Title} $(T)$.
    \item \emph{Decision} $(D)$ with \emph{Body} $(B)$.
    \item \emph{Context-Decision pairs} with \emph{Title-Body pairs}.
\end{itemize}

\rebuttal{While the core architecture remains consistent, the two tasks require distinct pipeline configurations for data extraction, and prompt engineering. These specific operational differences are detailed in subsequent sections.}

\section{\Approach: \ApproachFull}\label{sec:Approach}

\begin{figure*}[ht]
    \centering
    \includegraphics[width=1\textwidth]{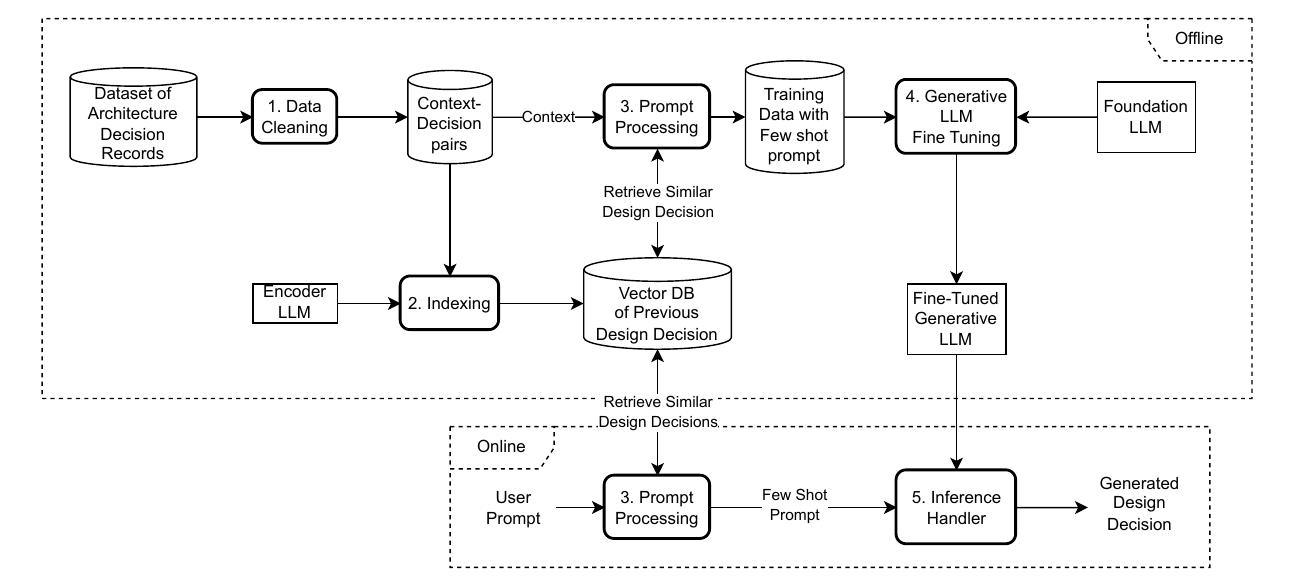}
    \caption{Detailed system diagram of \Approach~showing the five major processes: (1) data cleaning, (2) embedding and indexing into a vector database, (3) prompt processing for both offline training and online inference, (4) generative LLM \finetuning, and (5) inference handling to produce the final \Decision.}
    \label{fig:Approach}
\end{figure*}

This section presents \Approach, a \rebuttal{domain-specific} approach for generating ADR components. As established in Section \ref{sec:Overview}, we explain the approach using Task 1 (Context → Decision) as the illustrative example. All algorithms, system architecture, and processing steps apply identically to Task 2 (Title → Body) by substituting Title-Body pairs for Context-Decision pairs.
As mentioned in the previous section, \Approach~operates in 2 phases: an offline and an online phase. The details are provided in the following subsections.

\subsection{Offline phase}\label{sec:offline}

\begin{algorithm}[ht]
\caption{Generator Training} 
\label{alg:train_generator}
\small
\begin{algorithmic}[1]
\Procedure{TrainGenerator}{$\mathcal{ADR}$, $\text{LLM}_{\text{base}}$, $\text{LLM}_{\text{Encoder}}$, $k$}
    \State Data Cleaning
    \State $\mathcal{S} \gets \{\}$
    \For {$(C_i, D_i) \in \mathcal{ADR}$}
        \State Clean and standardize context $C_i$ and decision $D_i$
        \State Append $(C_i, D_i)$ to $\mathcal{S}$
    \EndFor

    \State $\mathcal{VDB} \gets \{\}$ 
    \For {each $(C_i, D_i) \in \mathcal{S}$}
        \State $\mathbf{v}_{C_i} \in \mathbb{R}^d \gets \text{LLM}_{\text{Encoder}}(C_i)$
        \State Store $(\mathbf{v}_{C_i}, (C_i, D_i))$ in vector database $\mathcal{VDB}$
    \EndFor
    \\
    
    \State Sample Context-Decision Pairs $\{(C', D')\}$ from $\mathcal{S}$
    
    \State $P \gets \textsc{PromptProcessing}(\mathcal{VDB}, C', \text{LLM}_{\text{Encoder}}, k)$
    
    \State $\text{LLM}_{\text{gen}} \gets \textsc{GenerativeLlmFineTuning}(\text{LLM}_{\text{base}}, P)$
    
    \State \textbf{Return:} $\text{LLM}_{\text{gen}}$
\EndProcedure
\end{algorithmic}
\end{algorithm}

In this phase, we execute the Generator Training procedure (as detailed in Algorithm \ref{alg:train_generator}) to fine-tune a foundational model for producing Decision based on given Contexts and similar retrieved examples. A visual representation of this process is provided in the offline section of Figure \ref{fig:Approach}.

As given in \textbf{Algorithm \ref{alg:train_generator}}, the process begins with data cleaning (lines 2–7), where the algorithm standardizes and organizes each ADR to extract Context ($C_i$) and Decision ($D_i$) pairs. This part is visually represented in process 1 of Figure \ref{fig:Approach}.
For each pair, it performs cleaning and standardization to ensure consistency across the dataset. This involves using string matching and regular expressions to extract and format the context and decision sections. Given that ADRs may have varying templates such as Nygard or MADR \footnote{\url{https://adr.github.io/adr-templates/}}, additional steps are taken to standardize the data, such as normalizing different section names (e.g., ``Context and Problem Statement" to ``Context"). This results in a cleaned and standardized dataset $\mathcal{S}$ of context-decision pairs (line 6), which is then ready for indexing and further use in training generative models.

In the next phase of \textbf{Indexing} (process 2 in Figure \ref{fig:Approach}), we create a vector database from the cleaned dataset $\mathcal{S}$ of context-decision pairs.
The context $C_i$ from each pair is converted into a high-dimensional vector embedding $v_{C_i}$ using an Encoder LLM (lines 8–12). These embeddings are stored in a vector database $\mathcal{VDB}$ along with their original context-decision pairs (line 11). In Figure \ref{fig:Approach}, this is shown as the `\textit{Vector DB of Previous Design Decision}' (refer Section \ref{sec:rag}). This indexed database is essential for the subsequent steps in prompt processing and fine-tuning, and also during inference.

Once the $\mathcal{VDB}$ has been constructed, the next step is to fine-tune the LLM. To initiate this process, it is necessary to generate a training prompt. The procedure begins by sampling a context-decision pair from the dataset (line 14), denoted as \( C'-D' \) . An illustrative example from a Mining Software Repositories study on use of ADRs in Open Source Projects \cite{adrs_data} of such a pair is provided in Figure \ref{fig:Sample_CD_pair}.

Algorithm 1 applies identically to Task 2, complete ADR generation. In lines 4-7, the algorithm extracts and cleans Title-Body pairs ($T_i, B_i$) instead of Context-Decision pairs. In lines 9-12, Titles ($T_i$) are embedded and stored in the $\mathcal{VDB}$ alongside their corresponding Bodies. In line 14, a Title-Body pair \( T'-B' \) is sampled for training. All other steps remain unchanged.

\begin{figure}[ht]
\centering
\begin{tcolorbox}[colback={highlight}, colframe={border}]
    \begin{rebuttalblue}
    \textbf{\{Context $C'$\}} \\
    {\scriptsize
    \par
The Office of Head Start is expanding data sharing across internal systems. To support this, TTA Hub would like to make documented APIs available to external systems in reliable, consistent, and secure ways.

\textbf{Options Considered}

To achieve this, two paths were considered.

\textbf{Expose existing internal APIs}

TTA Hub has a robust set of APIs already implemented to support the React frontend.

Pros:

\begin{itemize}
\item already implemented
\item well documented
\end{itemize}

Cons:

\begin{itemize}
\item API is highly coupled to TTA Hub UI needs, will become more difficult to change rapidly
\item authentication is based on sessions stored in browser cookies
\item authorization fully limited to existing TTA Hub roles
\end{itemize}

\textbf{Create a new API layer for external partners}

Creating a new API layer would de-couple the data needs from the TTA Hub UI needs.

Pros:

\begin{itemize}
\item existing API can continue to be optimized for UI considerations
\item new API can be more rigidly backwards-compatible
\item authentication schemes can be easily customized for each use
\item authorization can be scope-based and integrated into the full HSES OAuth implementation
\end{itemize}

Cons:

\begin{itemize}
\item two API endpoints will need to be maintained through data model changes
\end{itemize}
\par
    }
    \vspace{2mm}
    \textbf{\{Decision $D'$\}}\\
    {\scriptsize
    \par
A new API layer has been chosen as the better set of trade offs.

To distinguish between the API flavors, the external API will be namespaced under the /api/v1 route path.

\textbf{Authentication}\\
The initial API client also utilizes HSES as a single-sign-on solution, so API authentication will be done by passing an HSES token via the Authentication HTTP header. TTA Hub will then validate that token with HSES to verify the user.

\textbf{JSON Format}\\
To promote consistency, the external facing API will conform to the JSON:API schema as much as possible. The first endpoints will include objects in attributes that may eventually migrate to relationships. This trade-off has been made due to the exponential growth of required API endpoints when each object is represented in the relationships section.
\par
    }
    \end{rebuttalblue}
\end{tcolorbox}
\caption{Sample context–decision pair \( C'\text{-}D' \) from an ADR used as a training data in the \Approach~offline phase.}
\label{fig:Sample_CD_pair}
\end{figure}

\begin{algorithm}[ht]
\caption{Prompt Processing} 
\label{alg:prompt_processing_training}
\small
\begin{algorithmic}[1]
\Procedure{PromptProcessing}{$\mathcal{VDB}$, $C'$, $\text{LLM}_{\text{Encoder}}$, $k$}
    \State \( \mathbf{v}_{C'} \in \mathbb{R}^d \gets \text{LLM}_{\text{Encoder}}(C') \)
    \State \( \{(C_1, D_1), \dots,(C_k,D_k)\} \gets \text{RetrieveTopK}(\mathbf{v}_{C'}, \mathcal{VDB}) \)
    
    \State \( P = \{(C_1, D_1), (C_2, D_2), \dots, (C_k, D_k)\} + C' \)
    \State \textbf{Return:} \Fewshot~prompt $P$
\EndProcedure
\end{algorithmic}
\end{algorithm}

\textbf{Algorithm \ref{alg:prompt_processing_training}} then takes over to construct \fewshot~prompts for \finetuning~the generative LLM (line 16 in Algorithm \ref{alg:train_generator}). It corresponds to the process numbered 3 in Figure \ref{fig:Approach}. The process begins converting the given context to its embedding (line 2 in Algorithm \ref{alg:prompt_processing_training}) and retrieving the top-$k$ similar context-decision pairs from the vector database for each given context (line 3). Here, $k$ is a user-specified parameter, which is chosen on the basis of computational resources. 

Once the top-$k$ pairs are retrieved, the algorithm constructs a \fewshot~prompt $P$ for each context (line 4). This prompt includes the retrieved context-decision pairs as examples, followed by the given context for which a decision needs to be generated. The \fewshot~prompt provides the generative model with relevant examples, guiding it to produce accurate and contextually appropriate design decisions. This step is crucial for \finetuning~the generative model, as it helps the model to learn from similar past decisions and apply that knowledge to new contexts.

For example, let's take $k=2$ and assume 2 context-decisions pairs $(C_1,D_1)$, and $(C_2,D_2)$ from the source context $C'$ are retrieved. the constructed prompt would be as given in Figure \ref{fig:SampleInferencePrompt}.

Algorithm~2 applies identically to Task~2. In line~2, the input title $T'$ is embedded to obtain $\mathbf{v}_{T'}$. In line~3, the top-$k$ most similar title--body pairs $\{(T_1, B_1), \ldots, (T_k, B_k)\}$ are retrieved from the vector database (VDB). In line~4, these retrieved title--body pairs are combined with the input title $T'$ to form the few-shot prompt $P$. The overall algorithmic structure remains unchanged.


\begin{figure}[ht]
\centering
\begin{tcolorbox}[colback={highlight}, colframe={border}]
    \begin{rebuttalblue}
    \textbf{\{Retrieved Context 1 $C_1$\}} \\
    {\scriptsize
    \par
When we decided to start off with separate API and user-facing applications (see ADR 0004 0004-separate-api-and-user-facing-applications.md) we thought that we would need to build more than one service as part of this work, to meet different needs of different users. That may still be true in general, but from what we've learned in the Handover discovery it's looking more likely at the moment that the Handover phase of our work will involve expanding on Allocations rather than being an entirely separate journey and service.

We were also intending that the allocation manager would be a fairly minimal user-facing application, which would fetch all the data it needed from the allocation API and then render pages based directly on it.

We've been looking at how we should set up authentication on our two applications and have realised that that split of responsibilities between the allocation manager and API could lead to some confusions:

\begin{itemize}
\item Calling the Custody API from the allocation API means that the allocation API can be seen as sitting in two different layers depending on the context: either above the Custody API from the point of view of the allocation manager, but on the same level as the Custody API from the point of view of other services which need to use allocation data. This lack of clarity raises questions around which OAuth2 grant type to use for authentication on the allocation API.
\item Responsibility for access control may be split between the allocation manager and API. The user-facing application would need to know what the user should have access to in order to show the right options, but the API may also need to know what the user should have access to in order to fetch the right data from other APIs.
\end{itemize}

We still think there's value in having a separate allocation API, but it would be easier to understand if it were a more RESTful interface to the data it stores.
\par
    }
    \par\vspace{2mm}
    \textbf{\{Retrieved Decision 1 $D_1$\}}\\
    {\scriptsize
    \par
The allocation API will be a simpler interface to the data held in its database.

The allocation API will not call other APIs to return data from them to the allocation manager.

The allocation manager will call the allocation API and other APIs (Custody, Delius etc) directly.

The allocation manager will be entirely responsible for access control for its users.
\par
    }
    \par\vspace{2mm}
    \hrule
    \par\vspace{2mm}
    \textbf{\{Retrieved Context 2 $C_2$\}} \\
    {\scriptsize
    \par
As we move to a structure where products share services rather than being purely isolated systems, we are moving away from a model where the API Gateway acts solely as a pathway into the Sirius Backend API. Early integrations such as providing LPAs to \textbf{Use and LPA} or passing files from the \textbf{Digital Deputies Reporting Service} were thin layers of translation over the existing Sirius API.

The next phase of our API work is to move to a model where multiple products share and use an API that owns all its data, and encapsulates a need in the OPG domain. Our first example is the Use an LPA codes service, which provides data for both Sirius Case Management and the Use an LPA service, but is independent of both.

This is in parallel to work that is splitting Sirius into smaller services, with the expectation that the two streams of work will eventually converge as a set of isolated services that can be recomposed into new products, our ``Service Swarm''.
\par
    }
    \par\vspace{2mm}
    \textbf{\{Retrieved Decision 2 $D_2$\}}\\
    {\scriptsize
    \par
Shared services will sit behind the OPG Data API Gateway and follow the 0009-api-domain-structure.md. Shared services will be isolated and own a particular domain for our applications. That is, they own a particular business process or kind of data, for example the generation and expiry of access codes.

These services currently live within the boundary of the ``Sirius'' account, but are deployed independently.
\par
    }
    \end{rebuttalblue}
\end{tcolorbox}
\caption{Retrieved Context-Decision pairs for the example given in Figure \ref{fig:Sample_CD_pair}.}
\label{fig:Retrieved_CD_Prompt}
\end{figure}

\begin{algorithm}[ht]
\caption{Generative LLM Fine-tuning} 
\label{alg:fine_tuning}
\small
\begin{small}
\begin{algorithmic}[1]
\Procedure{GenerativeLlmFineTuning}{$\text{LLM}_{\text{base}}$, $P_{dataset}$}
    \For {each $P \in P_{dataset}$}
        \State $\hat{D'} \gets \text{LLM}_{\text{base}}(P)$
        \State $\mathcal{L} \gets \text{Loss}(\hat{D'}, D')$
        \State Update model parameters: 
        \[
        \theta_{\text{gen}} \leftarrow \theta_{\text{gen}} - \eta \nabla_{\theta_{\text{gen}}} \mathcal{L}
        \]
    \EndFor
    \State \textbf{Return:} Fine-tuned generative model $\text{LLM}_{\text{gen}}$
\EndProcedure
\end{algorithmic}
\end{small}
\end{algorithm}

Finally, the Generative LLM is \finetuned~using \textbf{Algorithm \ref{alg:fine_tuning}} titled \textbf{Generative LLM \Finetuning} (line 16 in Algorithm \ref{alg:train_generator}). This algorithm corresponds to the process numbered 4 in Figure \ref{fig:Approach}. It is designed to optimize a generative LLM to produce ADDs from the \fewshot~prompts $P$ constructed in the previous step.

The model generates a Decision $\hat{D'}$ for each prompt $P$ (line 3 in Algorithm \ref{alg:fine_tuning}). The difference between the generated decision $\hat{D'}$ and the actual decision $D'$ \rebuttal{is quantified using the standard cross-entropy loss over the target output tokens} which is then used to update the model's parameters through backpropagation (line 5).
This process is repeated for each prompt in the training set $P_{dataset}$ for multiple epochs, resulting in a \finetuned~LLM that can generate better ADDs based on a given contexts.

Algorithm 3 applies identically to Task 2. In line 3, the model generates a Body ($\hat{B}'$) instead of a Decision, and in line 4 the loss is computed against the ground truth Body ($B'$). Parameter updates in line 5 remain unchanged, yielding a model fine-tuned to generate complete ADR from title based prompts.

\subsection{Online phase}

\begin{algorithm}[H]
\caption{Inference} 
\label{alg:Inference}
\small
\begin{algorithmic}[1]
\Procedure{Inference}{$C, \text{LLM}_{\text{gen}}, \text{LLM}_{\text{Encoder}}, \mathcal{VDB}$, $k$}
    \State $P \gets \text{PromptProcessing}(\mathcal{VDB}, C, \text{LLM}_{\text{Encoder},} k)$
    
    \State $\hat{D} \gets LLM_{gen}(P)$ 
    \State \textbf{Return:} Final Decision $\hat{D}$
\EndProcedure
\end{algorithmic}
\end{algorithm}

In the online phase, inference is performed using the \finetuned~model as shown in Figure \ref{fig:Approach}. It gives us a decision $\hat{D}$ when prompted by a context $C$, as represented in \textbf{Algorithm \ref{alg:Inference}}. A sample context is given in Figure \ref{fig:SampleContextForInference}.


\begin{figure}[ht]
\centering
\begin{tcolorbox}[colback={highlight}, colframe={border}]
\begin{rebuttalblue}
    \textbf{\{Context $C$\}} \\
    {\scriptsize
    \par
In order to simplify the search infrastructure we are moving to using the publishing API as the source of data. This allows us to reduce the amount of code duplication across\textbackslash\{\}nprojects, as currently each publishing app has to implement a write interface into rummager.

We expect this will give us a number of improvements, including:

\begin{itemize}
\item rebuild the search index more easily
\end{itemize}

\begin{itemize}
\item reduce the load on the publishing API - rummager currently needs to retrieve tagging data from the publishing API when it indexes documents
\end{itemize}

The publishing API currently uses rabbit MQ to tell downstream apps that a new publishing event has occurred, with the key specifying the publishing app and the publishing event type.

Rummager currently processes `link` events from the rabbit MQ stream, this works by reading the message off the rabbitMQ stream, doing a small amount of manipulation and then creating\textbackslash\{\}na Sidekiq worker task to perform the data update.

When processing messages off the queue it is important that the process can provide the following features:

\begin{itemize}
\item Concurrency
\item Failure/Queue side tracking
\item Retry logic
\end{itemize}
\par
    }
\end{rebuttalblue}
\end{tcolorbox}
\caption{Sample Decision Context to illustrate \Approach 's online inference}
\label{fig:SampleContextForInference}
\end{figure}

The first step is \textbf{Prompt Processing} (line 2). Similar to the Prompt Processing in the offline phase, the VDB retrieves the most relevant context-decision $(C_i,D_i)$ pairs, where each context $C_i$ is similar to the provided context $C$. Let's assume the retrieved pairs are $(C_1,D_1)$ and $(C_2,D_2)$.
These context-decision pairs, along with the provided context \( C \), form a \fewshot~prompt as illustrated in Figure \ref{fig:SampleInferencePrompt}.

\begin{figure}[ht]
\centering
\begin{tcolorbox}[colback={highlight}, colframe={border}]
\begin{rebuttalblue}
    \textbf{\{instruction\}} \\
    {\scriptsize
    You are an expert software architect tasked with making decisions for Architectural Decision Records (ADRs). Given a context, provide a decision. The following examples illustrate the expected style.
    }\par
    \vspace*{2mm}
    
    \textbf{\{Retrieved Context 1 $C_1$\}} \\
    {\scriptsize
    \par
In [ADR 003 - perform popularity updating without using an index lock] docs/arch/adr-003-popularity-updating-without-index-locks.md we changed the process for bulk loading data into Rummager's search indexes to update the existing index in place, rather than creating a separate index, which had required locking the existing index until the new one was ready. The main reason we bulk load data is to update the popularity field every night, which affects every document in the search index.
\par
    }\par
    \vspace*{2mm}
    
    \textbf{\{Retrieved Decision 1 $D_1$\}}\\
    {\scriptsize
    \par
We decided against making any technical changes to the popularity update process now. Instead we reduced the number of sidekiq processes that index into elasticsearch at the same time, which reduced the impact on search performance,\textbackslash\{\}nwithout making any noticeable difference to the indexing time.
\par
    }\par
    \vspace*{5mm}
    
    \textbf{\{Retrieved Context 2 $C_2$\}} \\
    {\scriptsize
    \par
ElasticSearch (ES) is used as the search engine for:

\begin{itemize}
\item UI search for vacancies\textbackslash\{\}n- Querying vacancies to construct email job alerts based on a users criteria.
\end{itemize}

They both currently use a ruby ES client library to submit queries. Both use a base client class(es) that search and job alerts functionality extend from.

Current search has been identified as not giving optimal search results and job alerts are based on very simplistic search rules currently. Is there a better tool to use that is more easily configurable?

Currently indexing and maintaining ranking of search terms is an engineering task.
\par
    }\par
    \vspace*{2mm}
    
    \textbf{\{Retrieved Decision 2 $D_2$\}}\\
    {\scriptsize
    \par
Algolia was identified an alternative as it:

\begin{itemize}
\item has easier/improved 'out of the box' functionality yielding better search results - is up to 200x faster than ES - has a comprehensive dashboard UI that means rankings/weightings could potentially be managed by wider range of people - has a comprehensive toolset including UI components and libraries
\end{itemize}
\par
    }\par
    \vspace*{5mm}
    
    \textbf{\{instruction\}} \\
    {\scriptsize
    Make sure to give decisions that are similar to the ones above. Now provide a decision according to the context given below.
    }\par
    \vspace*{2mm}
    
    \textbf{\{Context $C$\}} \\
    {\scriptsize
    \par
In order to simplify the search infrastructure we are moving to using the publishing API as the source of data. This allows us to reduce the amount of code duplication across\textbackslash\{\}nprojects, as currently each publishing app has to implement a write interface into rummager.

We expect this will give us a number of improvements, including:

\begin{itemize}
\item rebuild the search index more easily
\end{itemize}

\begin{itemize}
\item reduce the load on the publishing API - rummager currently needs to retrieve tagging data from the publishing API when it indexes documents
\end{itemize}

The publishing API currently uses rabbit MQ to tell downstream apps that a new publishing event has occurred, with the key specifying the publishing app and the publishing event type.

Rummager currently processes `link` events from the rabbit MQ stream, this works by reading the message off the rabbitMQ stream, doing a small amount of manipulation and then creating\textbackslash\{\}na Sidekiq worker task to perform the data update.

When processing messages off the queue it is important that the process can provide the following features:

\begin{itemize}
\item Concurrency
\item Failure/Queue side tracking
\item Retry logic
\end{itemize}
\par
    }
\end{rebuttalblue}
\end{tcolorbox}
\caption{Few-shot inference prompt containing two retrieved context--decision pairs followed by the input context.}
\label{fig:SampleInferencePrompt}
\end{figure}

The \finetuned~model is now used to generate a Decision \(\hat{D}\) based on the \fewshot~prompt as shown in process numbered 5 in Figure \ref{fig:Approach}.
The generated \Decision~for the sample \Context~is given in Figure \ref{fig:SampleGeneratedDesignDecision}.

The online phase for Task 2 (Complete ADR Generation) operates identically. Given an input Title $T$, the system: (1) embeds the title to create $v_T$; (2) retrieves the top-$k$ most similar Title-Body pairs from the VDB; (3) constructs a few-shot prompt containing the retrieved examples and the input title; and (4) generates a complete ADR Body $\hat{B}$.


\begin{figure}[ht]
\centering
\begin{tcolorbox}[colback={highlight}, colframe={border}]
\begin{rebuttalblue}
    \textbf{\{Generated Decision $\hat{D}$\}} \\
    {\scriptsize
    \par
RabbitMQ has a number of downsides when it comes to implementing a worker. For example, it doesn't support concurrency, and the message needs to be acknowledged before the work is completed. This means that we can't have a single worker that can handle multiple messages, as we would need to acknowledge the message before the worker has finished processing it. This means that we would need to create a new worker for each message, which would be inefficient.
\par
    }
\end{rebuttalblue}
\end{tcolorbox}
\caption{Design Decision generated by the \Approach 's fine-tuned model for the sample context in Fig. \ref{fig:SampleContextForInference}}
\label{fig:SampleGeneratedDesignDecision}
\end{figure}

\section{Experiments and Evaluation}\label{sec:results}

\begin{figure}[ht]
    \centering
    \includegraphics[width=1\linewidth]{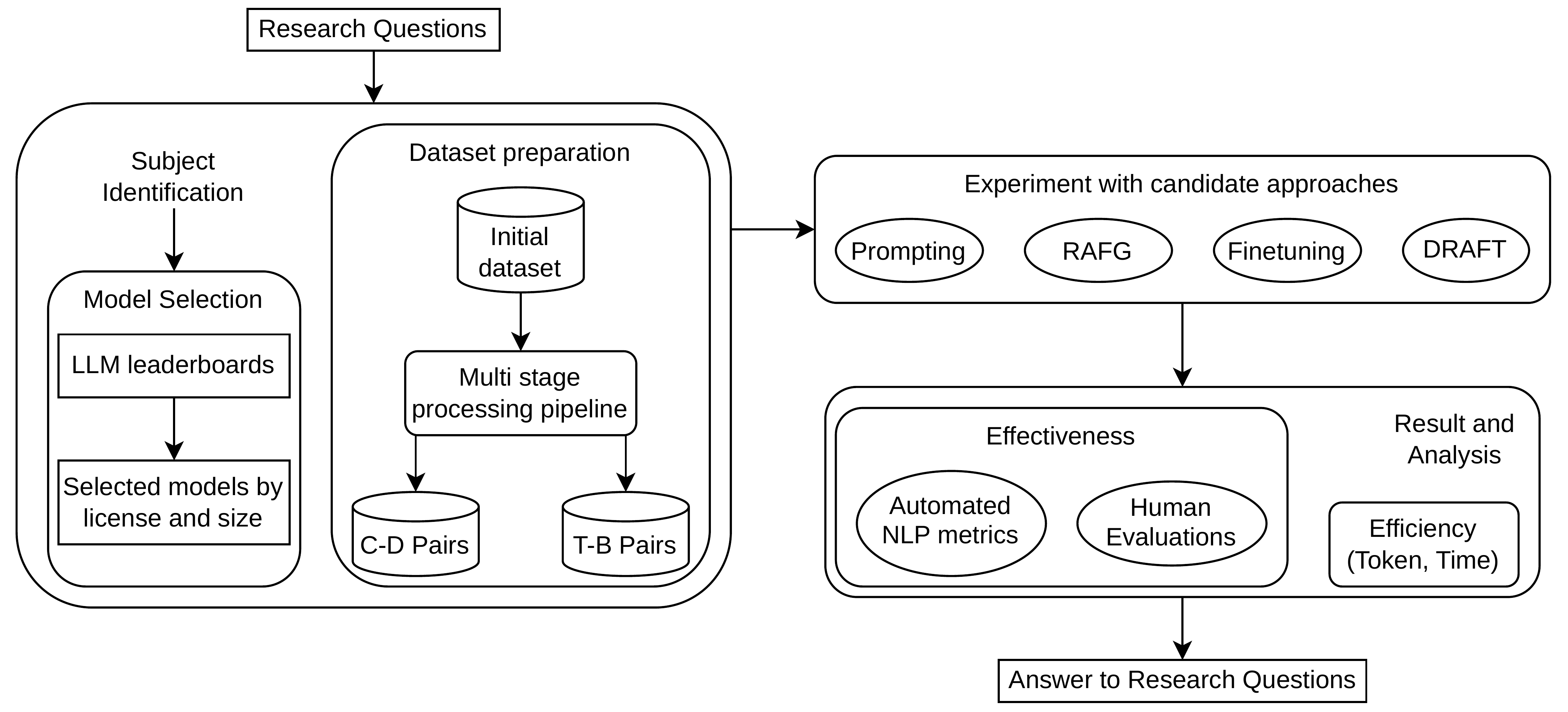}
    \caption{\rebuttal{Overview of the experimental procedure, encompassing research question formulation, subject identification (model selection and dataset preparation), experimentation with four candidate approaches (zero-shot prompting, \RAG, \finetuning, and \Approach), and result analysis using effectiveness and efficiency metrics.}}
    \label{fig:experiment}
\end{figure}

The objective of this evaluation is to assess both the effectiveness and efficiency of the proposed approach for both tasks: generating Design Decisions from Contexts (Task 1) and generating ADR Bodies from Titles (Task 2). To evaluate effectiveness, we measured the quality of the generated outputs by comparing them against the ground truth. To this end, we performed a series of controlled experiments using various LLMs for both tasks. Additionally, we separately evaluated the major components of \Approach, namely \RAFewshotG~and \finetuning, to analyze their individual contributions. This enabled us to gain a deeper understanding of each component's impact and to evaluate the overall effectiveness of \Approach~ in relation to these components. The experiments are designed to address the following key research questions, which guide the investigation of the strengths and limitations of our methodology. The experimental procedure involved formulating research questions, selecting the experimental subjects, conducting experiments using the defined approaches, evaluating the results, and addressing the research questions, as illustrated in Figure \ref{fig:experiment}.

\textbf{RQ1} \textit{How similar are the \Decision s (Task 1), and complete ADRs (Task 2) generated by existing approaches to those documented in real software repositories?}

This question aims to establish a performance baseline by evaluating how well commonly used LLM-based techniques (Prompting, \RAFewshotG~, and \finetuning) can replicate the quality and characteristics of human-written architectural documentation for both Task 1 (Context → Decision) and Task 2 (Title → Body). Understanding the capability of each method to produce outputs that align with real-world ADR practices is crucial for assessing their practical applicability. Similarity is measured using automated metrics that quantify lexical and semantic alignment with ground truth outputs from actual repositories, as detailed in Section~\ref{subsec:Evaluation}.

\textbf{RQ2} \textit{How does \Approach~compare to existing methods in generating \Decision s and complete ADRs that align with real repository documentation?}

This question evaluates whether our proposed approach (\Approach) generates outputs that are more similar to real-world architectural documentation compared to established techniques (Prompting, \RAFewshotG~, and \finetuning). It is important to assess whether combining \finetuning~and \RAFewshotG~delivers superior alignment with human-written ADRs for both the tasks, demonstrating practical value for AKM.

\textbf{RQ3} \textit{How does the efficiency of \Approach~compare to that of existing methods for generating \Decision s and complete ADRs?}

While quality is essential, practical deployment also depends on computational efficiency. This question investigates whether \Approach~introduces any significant overhead in terms of token usage and response time for both tasks, which are key considerations for time and cost-sensitive environments. Given the different computational characteristics of the two tasks (Task 1 with shorter inputs/outputs vs. Task 2 with longer ones), understanding efficiency trade-offs across both is crucial for practical adoption.

\begin{rebuttalblue}
RQ1 establishes the \textbf{baselines}, and RQ2 and compares \Approach~against the three baselines: zero-shot Prompting, RAFG, and Fine-tuning (Section~5.3). These baseline approaches are the standard ways of adapting a pre-trained LLM to a task, and RAFG and Fine-tuning are the two components of \Approach, so the comparison also isolates the contribution of each component.
\end{rebuttalblue}

\noindent All experiment source code and data are available on Zenodo\footnote{\url{https://zenodo.org/records/23438622}} and GitHub \footnote{\url{https://github.com/sa4s-serc/DRAFT}}.


\subsection{Dataset}\label{subsec:Dataset}

\begin{rebuttalblue}
We began by selecting the experimental subjects, namely the model and the dataset, as illustrated in Figure \ref{fig:experiment}. For our experiments, we used the ADR dataset released by Buchgeher et al. \cite{adrs_data}, which was collected from open-source GitHub repositories as part of their Mining Software Repositories (MSR) study. Buchgeher et al. conducted a systematic search across public GitHub repositories, combining automated filtering with manual validation and applying explicit inclusion and exclusion criteria to identify 921 repositories containing ADRs. They justified their choice of GitHub based on its position as the largest open-source hosting platform. The authors also acknowledged that, because their dataset is limited to open-source projects, the generalizability of their findings to professional or closed-source industrial software remains a threat to external validity.
\end{rebuttalblue}

We developed a comprehensive multi-stage pipeline to collect, process, and prepare a high-quality dataset of ADRs, as described below.

\medskip
\noindent \textbf{Aggregate Metadata}\\
The MSR study provided JSON files containing repository metadata including URLs and ADR locations. We processed the JSON files to extract repository details and ADR paths, and also counted template types (such as Nygard and MADR \cite{madr}). It resulted in a \textit{csv} listing all ADR locations.

\medskip
\noindent \textbf{Scrape ADR Files}\\
Using the \textit{csv}, we downloaded the actual ADR markdown files from GitHub repositories. The scraping process implemented a robust fallback mechanism for branch detection, attempting 'main', then 'master', and finally querying the GitHub API for available branches when needed. All downloaded files were organized into repository-specific folders.

\medskip
\noindent \textbf{Dataset Creation}\\
The raw markdown files were transformed into a structured dataset. We extracted both Context-Decision (C-D) pairs and Title-Body (T-B) pairs from the cleaned ADRs to support both experimental tasks. Recognizing the variability in ADR formats, the extraction process used systematic techniques including string matching and regular expressions to identify and extract the relevant components.
The process included:
\begin{itemize}
    \item \textit{Title Extraction:} Titles were identified using YAML frontmatter, heading parsing, or filename fallback
    \item \textit{Section Parsing:} Key sections (Context and Decision) were extracted by searching for common heading patterns
    \item \textit{Context Fallback:} When explicit "Context" sections were absent, all content preceding the "Decision" section was designated as context
    \item ADRs where decisions could not be extracted were excluded
\end{itemize}
This stage produced a JSONL file where each line (datapoint) represents one ADR with fields including Title, Body, Context, Decision, and token counts. Hence we now had the C-D and the T-B pairs.

\medskip
\noindent \textbf{Filtering}\\
To ensure data quality, we applied rigorous filtering criteria:
\begin{itemize}
    \item \textit{Token Count:} Removed ADRs with bodies, contexts, or decisions outside acceptable ranges (e.g., body between 10-1000 tokens)
    \item \textit{Language:} Filtered non-English ADRs by requiring at least 90\% printable ASCII characters
    \item \textit{URL Density:} Excluded ADRs with excessive URLs (>20\% of tokens in context or decision)
\end{itemize}

\medskip
\noindent \textbf{De-duplication}\\
Near-identical ADRs were removed using semantic similarity:
\begin{itemize}
    \item ADR bodies were converted to vector embeddings using the sentence-transformers library with the 'all-MiniLM-L6-v2' model
    \item Cosine similarity was calculated between all pairs of embeddings
    \item ADRs with similarity $\ge$ 0.98 were marked as duplicates and removed
    \item A log file tracked which ADRs were removed and their duplicates
\end{itemize}

\medskip
\noindent \textbf{Split}\\
After all processing stages, the final dataset consisted of 4,344 ADRs containing both C–D and T–B pairs. The dataset was randomly shuffled to ensure unbiased partitioning and subsequently split into training (70\%), validation (10\%), and test (20\%) sets, comprising 3,040, 434, and 870 ADRs, respectively.

This split was applied consistently to maintain the same ADRs in each set across both tasks, ensuring fair comparison. The training set was used for fine-tuning and \Approach -ing models, the validation set for model selection based on lowest validation loss, and the test set for final evaluation.

\rebuttal{To ensure rigorous evaluation and prevent data leakage, \Approach~utilizes strictly disjoint datasets across its phases. During the offline phase, the Vector Database (VDB) is constructed and the model is fine-tuned exclusively using the training set. In contrast, the online inference phase operates on a separate VDB constructed solely from the test set. Furthermore, in both phases, a post-retrieval check is performed to ensure that the query datapoint itself is not retrieved (because ideally it would be the closest match). This demonstrates the model's ability to generalize to new architectural contexts rather than merely memorizing the training database.}

\medskip
\noindent \textbf{Analysis}\\
We analyzed ADR component lengths using the tiktoken library. The median Context length was 56 tokens (approximately 45 words, assuming 5 tokens is roughly equal to 4 words), while the median Decision length was 73 tokens, indicating their typically concise nature. For Title-Body pairs, the median Title length was 8 tokens and the median Body length was 275 tokens, reflecting the more comprehensive nature of complete ADR documentation. These ADRs, authored by software architects in real projects, serve as ground truth for both tasks.
\subsection{LLM Selection}\label{subsec:LLMSelection}
\begin{small}
\begin{table}[ht]
\centering
\begin{tabular}{l|c|c|c|c}
    \toprule
    \textbf{Model Type} & \textbf{Provider} & \textbf{Model Name} & \textbf{Size} & \textbf{Availability}\\
    \midrule
    \multirow{4}{0.18\textwidth}{Generative Model} & OpenAI & gpt-5-high & unknown & proprietary\\
    & Google & Gemini-2.5-flash & unknown & proprietary\\
    & Alibaba & Qwen3-30b-a3b-instruct-2507 & 30B & open source\\
    & Google & Gemma-3-4b-it & 4B & open source\\
    \midrule
    \multirow{2}{*}{Embedding Model} & OpenAI & text-embedding-3-large & hidden & proprietary\\
    & Google & gemini-embedding-001 & hidden & proprietary\\
    & Alibaba & Qwen3-Embedding-8B & 8B & open source\\
    \bottomrule
\end{tabular}
\caption{LLMs used in this study, listing provider, model name, size, and availability, for generative and embedding models.}
\label{tab:LLM}
\end{table}
\end{small}

\begin{table}[ht]
\centering
\begin{tabular}{l|c}
    \toprule
    \textbf{Generative Model} & \textbf{Embedding Model}\\
    \midrule
    Gpt-5-high & text-embedding-3-large\\
    Gemini-2.5-flash & gemini-embedding-001\\
    Qwen3-30b-a3b-instruct-2507 & Qwen3-Embedding-8B\\
    Gemma-3-4b-it & Qwen3-Embedding-8B\\
    \bottomrule
\end{tabular}
\caption{Pairing of generative and embedding models}
\label{tab:LLMPairing}
\end{table}

The availability of numerous generative LLMs from different organizations makes model selection a critical step. We selected models based on established leaderboards, size, and license.

\medskip
\noindent \textbf{Generative Model Selection}

We referred to the LM Arena text rankings \cite{chiang2024chatbotarenaopenplatform}\footnote{\url{https://arena.ai/}}, a well-accepted platform for human preference-based LLM evaluation, and selected top performing models as of September 8, 2025. We recognize that the LM Arena leaderboard is fast-evolving due to fast-paced innovation in this field, but we believe that the models used in this study provide a good representation of LLMs in general. We included both proprietary and open-source models to test how well DRAFT adapts to various architectures and sizes.

Table \ref{tab:LLM} lists all the selected models.
We selected GPT-5-high and Gemini-2.5-Flash as proprietary models, representing top-tier commercial LLMs. These models deliver state-of-the-art performance but come with vendor lock-in and data privacy concerns.
For organizations seeking to avoid third-party data sharing, we offer open-source models that can be deployed on-premises. The Qwen3-30B-A3B-Instruct-2507, a 30 billion parameter model, provides high performance for advanced computational setups, while Gemma-3-4B-it, with 4 billion parameters, is tailored for organizations with more limited hardware resources, allowing fine-tuning on modest systems.

Although the exact parameter counts of GPT-5-high and Gemini-2.5-flash are not publicly disclosed, this work makes assumptions based on commonly reported estimates in the literature and community discussions. Specifically, we assume that GPT-5-high represents the largest model in this study, with a parameter scale on the order of trillions, while Gemini-2.5-flash is assumed to be the next largest model, with a parameter scale on the order of hundreds of billions.

\medskip
\noindent \textbf{Embedding Model Selection}

For the retrieval components required by \RAG~and \Approach~approaches, we selected embedding models following a two-pronged strategy.

\textbf{Open Source Embedding Model:} We selected Qwen3-Embedding-8B, which ranks at the top of the MTEB Retrieval leaderboard (https://huggingface.co/spaces/mteb/leaderboard) as of September 8, 2025. This model was paired with all open source generative models (Qwen3-30B and Gemma-3-4B) to support effective retrieval for open-source deployments.

Organizations using proprietary generative models often face vendor lock-in, as they rely on the same provider for both generative and embedding models to maintain ecosystem compatibility. Accordingly, we selected the embedding models recommended by each LLM provider: text-embedding-3-large for GPT-5-high (OpenAI) and gemini-embedding-001 for Gemini-2.5-Flash (Google), representing the best available options from each vendor at the time of selection, as detailed in table \ref{tab:LLM}.

\medskip
\noindent \textbf{Pairing}

The pairing of generative and embedding models used throughout our experiments follows a vendor-alignment strategy, as shown in Table \ref{tab:LLMPairing}. For proprietary models, we paired each generative model with its corresponding embedding model from the same provider. For open-source models, both Qwen3-30B-A3B-Instruct-2507 and Gemma-3-4B-it were paired with Qwen3-Embedding-8B, the top-performing open-source embedding model from the MTEB leaderboard. This pairing strategy ensures that organizations can deploy either fully proprietary or fully open-source solutions while maintaining state-of-the-art retrieval capabilities.
\subsection{Experimental Candidates} \label{subsec:expCandidates}

The candidate approaches chosen for the experimentation as Prompting, \RAFewshotG, \finetuning~and \Approach~as illustrated in Figure \ref{fig:experiment}. All approaches were evaluated on both Task 1 (Context → Decision) and Task 2 (Title → Body). The system prompts used in the experiments can be found along with the corresponding code in the replication package.

\medskip
\noindent \textbf{Prompting}

To evaluate zero-shot prompting for both tasks, we provided each model with the appropriate input (Context for Task 1, Title for Task 2) accompanied by a suitable system prompt, expecting it to generate the corresponding output (Decision for Task 1, Body for Task 2). This setup reflects a straightforward interaction with an LLM, similar to typical usage by individuals without specialized knowledge in \finetuning~or configuration. The resulting outputs from this experiment establish the baseline for evaluating and comparing \Approach~with the other approaches across both tasks.

\medskip
\noindent \textbf{\RAFewshotG} 

This experiment evaluates the effectiveness of \RAFewshotG~in improving LLM performance for generating Architectural Artifacts across both tasks. The objective was to determine whether retrieval-augmented few-shot prompting could enhance the model's ability to produce more relevant and contextually appropriate outputs compared to baseline.

To implement this approach, we created two separate VDBs: one containing C-D pairs for Task 1 and another containing T-B pairs for Task 2. For Task 1, each context-decision pair was represented by a vector embedding of the context generated by an encoder model. When given a new input context, its embedding was computed by the encoder LLM, and the top 2 most similar contexts and their corresponding decisions were retrieved (as explained in Section \ref{sec:rag}). For Task 2, the process was identical but using title-body pairs, where titles were embedded and used for retrieval. These examples were then used to construct \fewshot~prompts, helping the model generate outputs based on patterns observed in similar inputs. The embedding models and paired LLMs used in this process are listed in Table \ref{tab:LLMPairing}.

\medskip
\noindent \textbf{\Finetuning}

In this experiment, we assess the impact of \finetuning~a foundational model for both the tasks. The dataset was divided into a train-validation-test split as described in Section \ref{subsec:Dataset}. Models were then \finetuned~separately for Task 1 (C-D pairs) and Task 2 (T-B pairs).

Models were \finetuned~on the training set for a maximum of 5 epochs, with a checkpoint saved at the end of each epoch. The final model was selected based on the checkpoint with the lowest validation loss, after which the outputs were generated for the test set, and performance was evaluated using the predefined metrics. We \finetuned~for 5 epochs as all validation losses converged (validation loss dropped to its lowest point and then increased again) within 5 epochs.

\medskip
\noindent \textbf{\Approach}

In this experiment, we assess the impact of \Approach-ing foundational models to generate outputs for both tasks. The initial dataset was divided into a train-validation-test split as described in Section \ref{subsec:Dataset}. Then the dataset was processed to form few-shot prompts as described in Section \ref{sec:offline}, separately for each task. Models were \Approach-ed on the training set for a maximum of 5 epochs, with a checkpoint saved at the end of each epoch. The final model was selected based on the checkpoint with the lowest validation loss after which the artifacts were generated for the test set, and performance was evaluated using the predefined metrics similar to \finetuning~approach. We \Approach-ed for 5 epochs as all validation losses converged within 5 epochs.

\medskip
\noindent \textbf{Common setup for \finetuning~and \Approach}

\begin{figure}[ht]
  \centering
  \includegraphics[page=1,width=\linewidth]{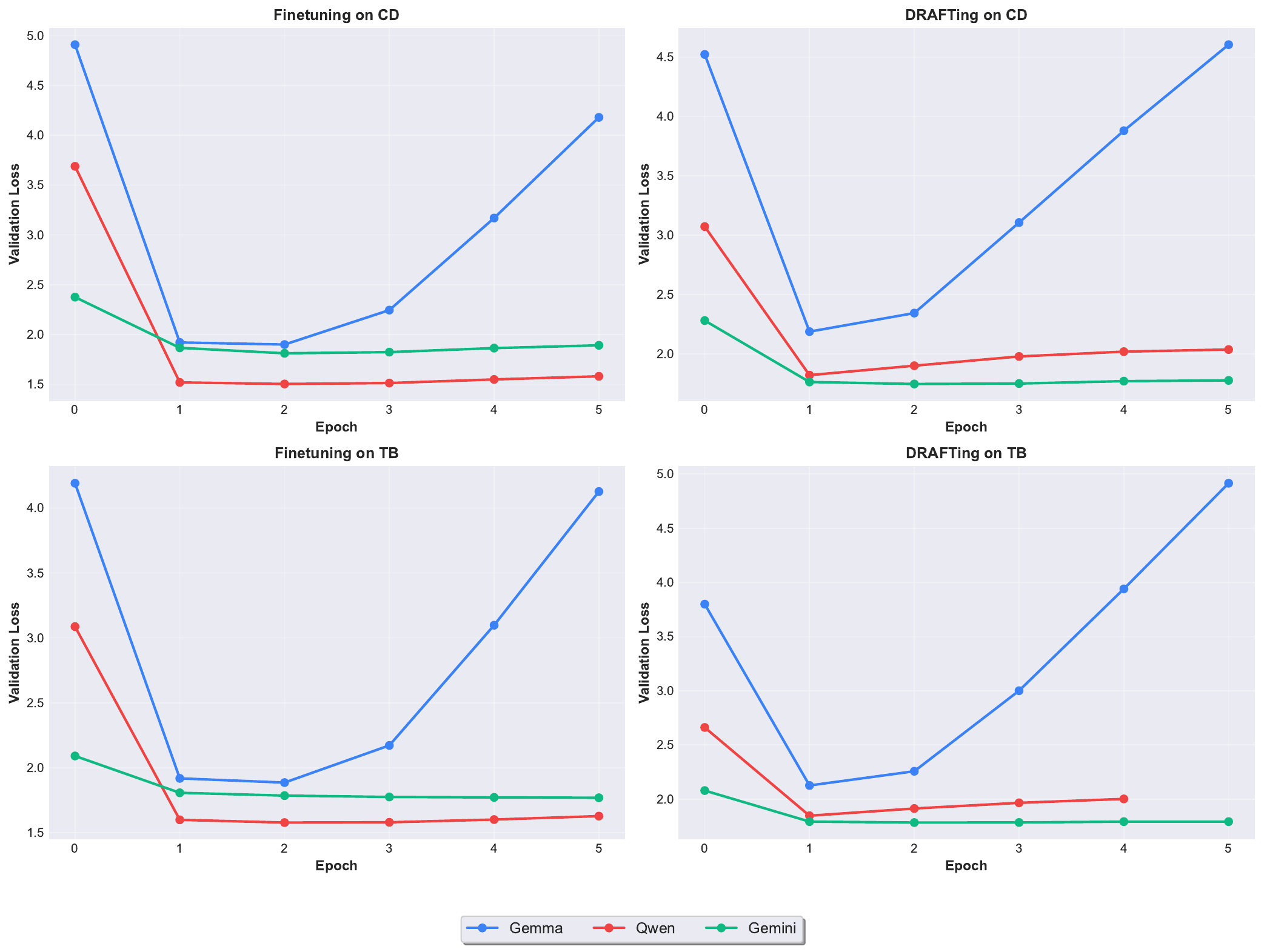}
  \caption{Validation loss curves across 5 epochs for 3 LLMs on \finetuning~and \Approach~for both tasks}
  \label{fig:Validation_Loss}
\end{figure}

Three models were \textit{training} (training refers to both \finetuning~and \Approach~in this Section) for both tasks: two open-source models (Qwen3-30B-A3B-Instruct-2507 and Gemma-3-4B-it) and one proprietary model (Gemini-2.5-Flash). 
Gemma underwent a full-parameter training, whereas Qwen was trained using Low-Rank Adaptation (LoRA) (refer Section \ref{sec:Finetuning}). Training was conducted on a server with Nvidia A6000 GPU \footnote{\url{https://www.nvidia.com/en-in/products/workstations/rtx-a6000/}} having 48GB V-ram, and 128GB Ram. Gemini was trained using it's training API.

To select the best model for inference, we chose the one with the lowest validation loss as a lower validation loss shows that the model was able to generalize well to unseen data and is more likely to perform well during inference. The validation loss curves for \finetuning~and \Approach~are shown in Figure \ref{fig:Validation_Loss}. The validation shows results from the 0th epoch, i.e. the untrained model with no changes, till the final epoch of training.

Please note that in this Section, \textit{training} refers to both \finetuning~and \Approach. GPT-5 was not \finetuned~or \Approach-ed as it was not available for \finetuning~at the time of the study, and was only evaluated using prompting and RAFG approaches.

\subsection{Evaluation Setup}\label{subsec:Evaluation}

\begin{table}[ht]
\centering
\begin{tabular}{>{\centering\arraybackslash}p{0.14\textwidth}|
>{\centering\arraybackslash}p{0.12\textwidth}|
>{\centering\arraybackslash}p{0.07\textwidth}
>{\centering\arraybackslash}p{0.06\textwidth}
>{\centering\arraybackslash}p{0.07\textwidth}
>{\centering\arraybackslash}p{0.11\textwidth}|
>{\centering\arraybackslash}p{0.06\textwidth}
>{\centering\arraybackslash}p{0.09\textwidth}
>{\centering\arraybackslash}p{0.06\textwidth}
}
    \toprule
    \textbf{Model} & \textbf{Approach} & \multicolumn{2}{c}{\textbf{ROUGE}} & \textbf{BLEU}  & \textbf{METEOR} & \multicolumn{3}{c}{\textbf{BERTScore}}  \\
    \cmidrule(lr){3-4}
    \cmidrule(lr){7-9}
    & & 1 & Lsum & & & Precision & Recall & F1  \\
    \midrule
    \multirow{4}{0.18\textwidth}{Gemma-3-4b} & Prompting & \cellcolor{worst}0.135 & \cellcolor{worst}0.122 & \cellcolor{worst}0.009 & \cellcolor{best}0.165 & \cellcolor{worst}0.780 & \cellcolor{worst}0.831 & \cellcolor{worst}0.805 \\
        & Finetuning & 0.186 & 0.164 & \cellcolor{best}0.032 & 0.164 & 0.833 & 0.836 & 0.834  \\
        & RAFG & \cellcolor{best}0.191 & \cellcolor{best}0.165 & 0.020 & \cellcolor{worst}0.159 & \cellcolor{best}0.849 & \cellcolor{best}0.838  & \cellcolor{best}0.843  \\
        & DRAFT & 0.179 & 0.159 & 0.031 & 0.162 & 0.828 & 0.835 & 0.831 \\
    \midrule
    \multirow{4}{0.18\textwidth}{Qwen-3-30B} & Prompting & \cellcolor{worst}0.159 & \cellcolor{worst}0.131 & \cellcolor{worst}0.014 & 0.184 & \cellcolor{worst}0.807 & 0.838 & \cellcolor{worst}0.822  \\
        & Finetuning & 0.210 & 0.188 & 0.033 & \cellcolor{worst}0.162 & 0.859 & 0.837 & 0.847  \\
        & RAFG & 0.183 & 0.148 & 0.020 & \cellcolor{best}0.190 & 0.859 & 0.837 & 0.847 \\
        & DRAFT & \cellcolor{best}0.213 & \cellcolor{best}0.191 & \cellcolor{best}0.041 & 0.167 & \cellcolor{best}0.863 & 0.838 & \cellcolor{best}0.850  \\
    \midrule
    \multirow{4}{0.18\textwidth}{gemini-2.5} & Prompting & 0.162 & 0.133 & \cellcolor{worst}0.015 & 0.158 & \cellcolor{worst}0.825 & 0.835  & 0.829  \\
        & Finetuning & \cellcolor{worst}0.125 & \cellcolor{worst}0.114 & 0.022 & \cellcolor{worst}0.103 & 0.835 & \cellcolor{worst}0.820 & \cellcolor{worst}0.827  \\
        & RAFG & 0.176 & 0.147 & 0.019 & 0.172 & 0.830 & 0.838 & 0.834  \\
        & DRAFT & \cellcolor{best}0.199 & \cellcolor{best}0.177 & \cellcolor{best}0.040 & \cellcolor{best}0.178 & \cellcolor{best}0.837 & \cellcolor{best}0.841 & \cellcolor{best}0.839  \\
    \midrule
    \multirow{2}{0.18\textwidth}{gpt-5} & Prompting & 0.138 & 0.123 & 0.008 & 0.164 & 0.789 & 0.836  & 0.812  \\
        & RAFG & 0.146 & 0.130 & 0.009  & 0.169 & 0.794 & 0.838 & 0.815  \\
    \bottomrule
\end{tabular}
\caption{Effectiveness results for Task 1 (Context → Decision) across all model–approach combinations, reporting ROUGE-1, ROUGE-Lsum, BLEU, METEOR, and BERTScore (precision, recall, F1) on the held-out test set of 870 ADRs.}
\label{tab:results_cd}
\end{table}

\begin{table}[ht]
\centering
\begin{tabular}{>{\centering\arraybackslash}p{0.14\textwidth}|
>{\centering\arraybackslash}p{0.12\textwidth}|
>{\centering\arraybackslash}p{0.07\textwidth}
>{\centering\arraybackslash}p{0.06\textwidth}
>{\centering\arraybackslash}p{0.07\textwidth}
>{\centering\arraybackslash}p{0.11\textwidth}|
>{\centering\arraybackslash}p{0.06\textwidth}
>{\centering\arraybackslash}p{0.09\textwidth}
>{\centering\arraybackslash}p{0.06\textwidth}
}
    \toprule
        \textbf{Model} & \textbf{Approach} & \multicolumn{2}{c}{\textbf{ROUGE}} & \textbf{BLEU}  & \textbf{METEOR} & \multicolumn{3}{c}{\textbf{BERTScore}}  \\
    \cmidrule(lr){3-4}
    \cmidrule(lr){7-9}
    & & 1 & Lsum & & & Precision & Recall & F1  \\
    \midrule
    \multirow{4}{0.18\textwidth}{Gemma-3-4b} & Prompting & \cellcolor{worst}0.221 & \cellcolor{worst}0.208 & \cellcolor{worst}0.013 & 0.222 & \cellcolor{worst}0.808 & \cellcolor{worst}0.821 & \cellcolor{worst}0.814 \\
        & Finetuning & \cellcolor{best}0.269 & \cellcolor{best}0.255 & \cellcolor{best}0.062 & 0.245 & 0.831 & 0.829 & 0.830  \\
        & RAFG & 0.247 & 0.233 & 0.036 & \cellcolor{best}0.255 & 0.814 & \cellcolor{best}0.831 & 0.822  \\
        & DRAFT & 0.263 & 0.248 & 0.051 & \cellcolor{worst}0.221 & \cellcolor{best}0.837 & 0.828 & \cellcolor{best}0.832 \\
    \midrule
    \multirow{4}{0.18\textwidth}{Qwen-3-30B} & Prompting & 0.242 & 0.227 & \cellcolor{worst}0.026 & 0.226 & 0.809 & 0.825 & 0.817  \\
        & Finetuning & \cellcolor{worst}0.237 & \cellcolor{worst}0.223 & \cellcolor{best}0.050 & \cellcolor{worst}0.192 & 0.850 & 0.822 & 0.835  \\
        & RAFG & 0.242 & 0.227 & 0.027 & 0.226 & 0.809 & 0.825 & 0.817 \\
        & DRAFT & \cellcolor{best}0.251 & \cellcolor{best}0.236 & 0.035 & 0.194 & \cellcolor{best}0.858 & 0.822 & \cellcolor{best}0.839  \\
    \midrule
    \multirow{4}{0.18\textwidth}{gemini-2.5} & Prompting & \cellcolor{worst}0.145 & \cellcolor{worst}0.135 & \cellcolor{worst}0.009 & \cellcolor{worst}0.118 & 0.832 & \cellcolor{worst}0.800  & 0.815  \\
        & Finetuning & 0.182 & 0.171 & 0.040 & 0.166 & 0.829 & 0.812 & 0.820  \\
        & RAFG & 0.158 & 0.148 & 0.032 & 0.157 & \cellcolor{worst}0.821 & 0.807 & \cellcolor{worst}0.813  \\
        & DRAFT & \cellcolor{best}0.213 & \cellcolor{best}0.199 & \cellcolor{best}0.041 & \cellcolor{best}0.185 & \cellcolor{best}0.844 & \cellcolor{best}0.819 & \cellcolor{best}0.830  \\
    \midrule
    \multirow{2}{0.18\textwidth}{gpt-5} & Prompting & 0.198 & 0.183 & 0.012 & 0.155 & 0.801 & 0.806  & 0.803  \\
        & RAFG & 0.204 & 0.188 & 0.013  & 0.153 & 0.805 & 0.807 & 0.806  \\
    \bottomrule
\end{tabular}
\caption{Effectiveness results for Task 2 (Title → Body) across all model–approach combinations, reporting ROUGE-1, ROUGE-Lsum, BLEU, METEOR, and BERTScore (precision, recall, F1) on the held-out test set of 870 ADRs.}
\label{tab:results_tb}
\end{table}


\noindent \textbf{Automated Metrics}

The next phase of the experiment involved the evaluation process, as depicted in Figure \ref{fig:experiment}. Evaluation of text generation often relies on a combination of metrics rather than a single metric. In line with this practice, our evaluation incorporates ROUGE-1 \cite{ROUGE}, BLEU score \cite{BLEU}, METEOR \cite{METEOR}, and BERTScore \cite{zhang2020bertscoreevaluatingtextgeneration}.

\textbf{ROUGE} (Recall-Oriented Understudy for Gisting Evaluation) \cite{ROUGE} is a set of metrics used to evaluate the quality of machine-generated summaries. We are using ROUGE-1 to measure the overlap of unigrams (individual words) between the system-generated text and the reference text.

\textbf{BLEU} (Bilingual Evaluation Understudy) \cite{BLEU} score is a precision metric used to evaluate the quality of machine-translated text.

\textbf{METEOR} (Metric for Evaluation of Translation with Explicit ORdering) \cite{METEOR} is a stronger metric used for evaluating machine-generated text, particularly in the context of machine translation.

\textbf{BERTScore} \cite{zhang2020bertscoreevaluatingtextgeneration} is an automatic evaluation metric used to assess the quality of text generation. It leverages pre-trained contextual embeddings from BERT (Bidirectional Encoder Representations from Transformers) \cite{devlin2019bertpretrainingdeepbidirectional} and measures the similarity between words in candidate and reference sentences using cosine similarity. BERTScore captures semantic similarity between texts and has been shown to correlate well with human judgment in evaluating text generation outputs \cite{zhang2020bertscoreevaluatingtextgeneration}. Hence, we use it as the primary metric in this study.


\begin{figure}[ht]
  \centering
  \includegraphics[page=1,width=\linewidth]{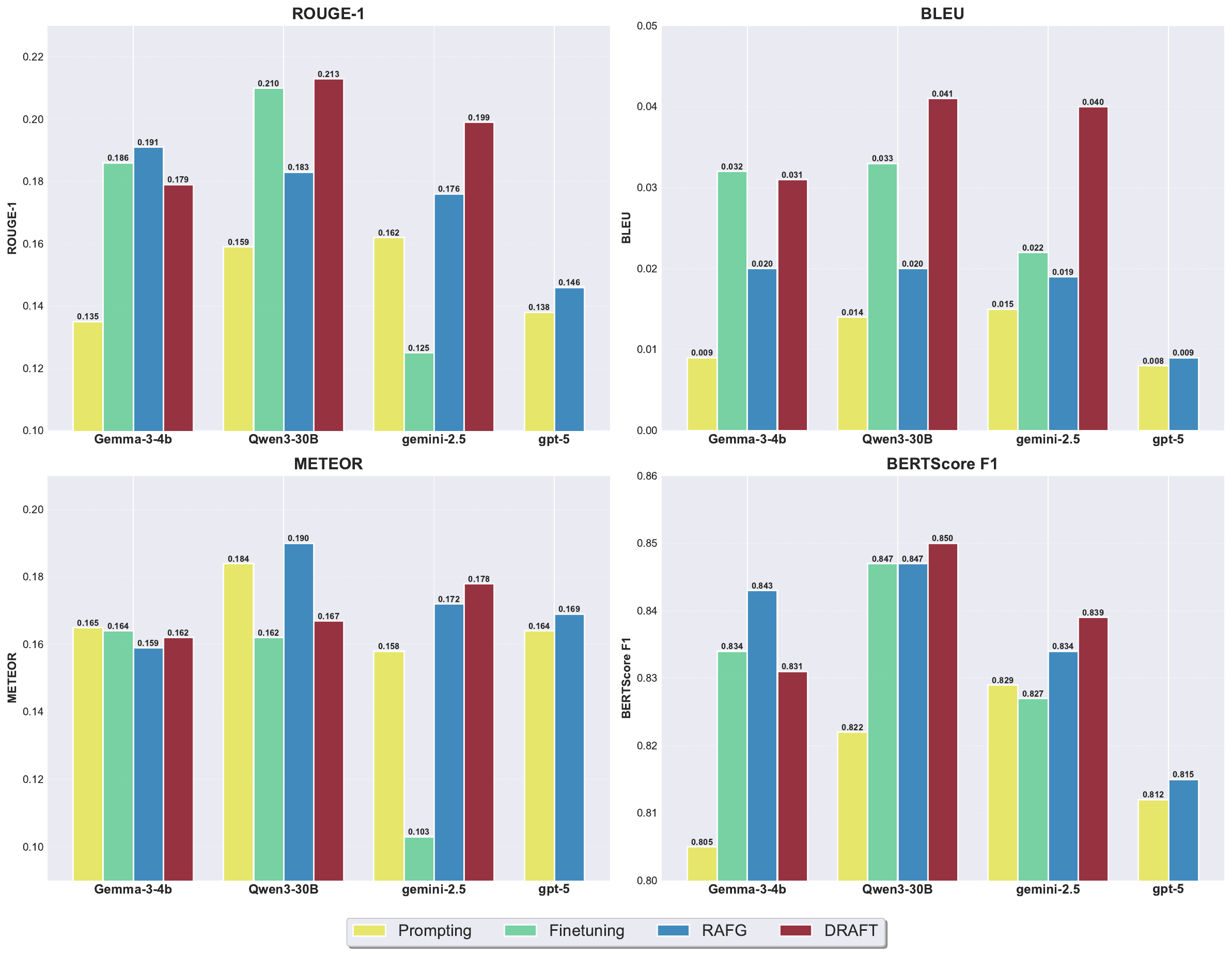}
  \caption{Comparison of all four approaches (Prompting, \Finetuning, \RAG, and \Approach) across 4 LLMs on Task 1 (Context → Decision), measured by ROUGE-1, BLEU, METEOR, and BERTScore F1.}
  \label{fig:CD_Results}
\end{figure}

\begin{figure}[ht]
  \centering
  \includegraphics[page=1,width=\linewidth]{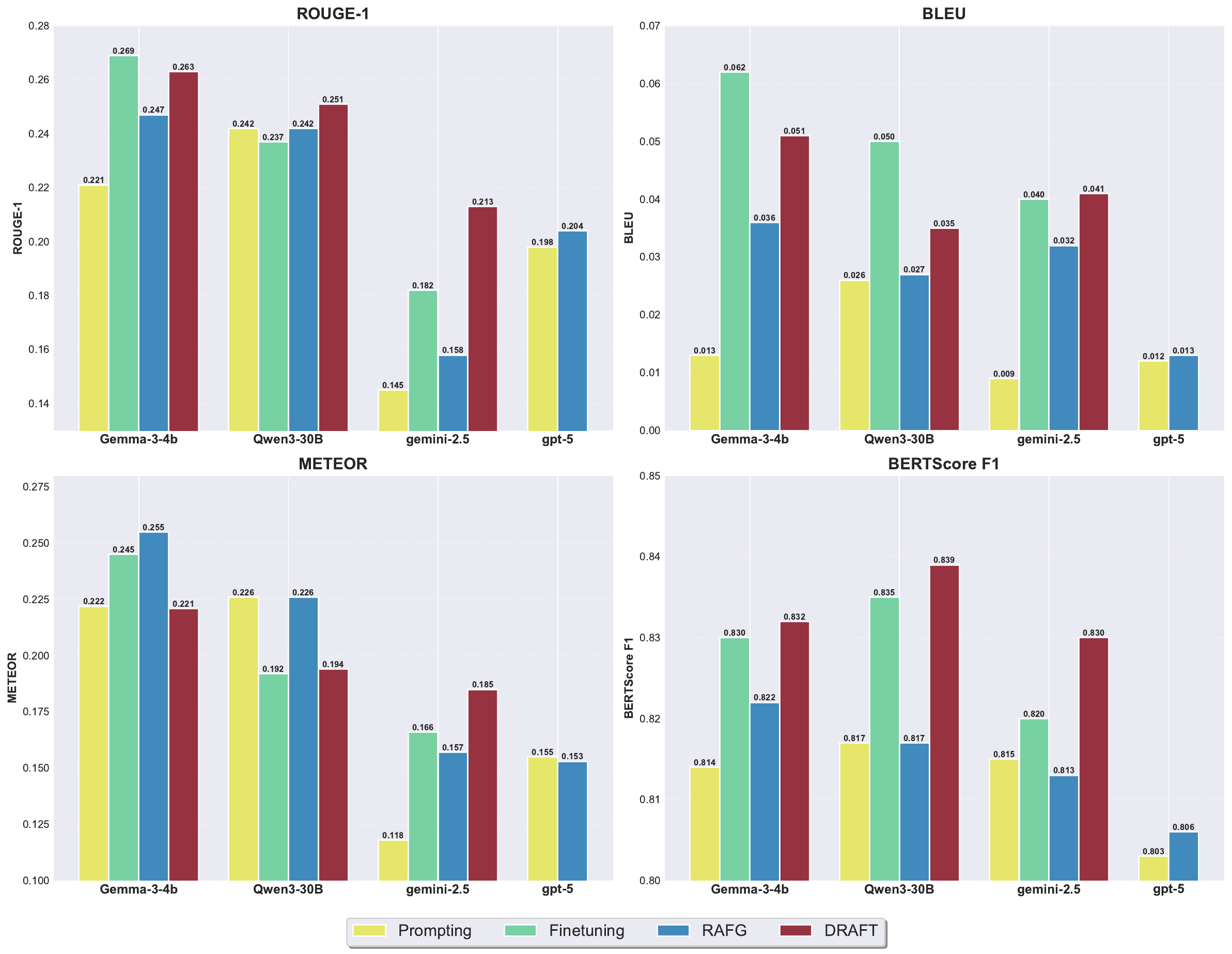}
  \caption{ Comparison of all four approaches (Prompting, \Finetuning, \RAG, and \Approach) across four LLMs on Task 2 (Title → Body), measured by ROUGE-1, BLEU, METEOR, and BERTScore F1.}
  \label{fig:TB_Results}
\end{figure}

\begin{rebuttalblue}

\medskip
\noindent \textbf{Human Evaluations}


To complement the automated metrics, we conducted a blinded human evaluation to assess the quality of \Decision~and ADR generation across the four approaches: Prompting, \Finetuning, \RAG, and \Approach. The evaluation consisted of two parts, both covering both tasks. Sample sizes for both parts were determined using the standard sample size formula with finite population correction, assuming a population proportion of 50\%.

\textbf{Author study:} 64 ADR instances were sampled from the 870 test instances, corresponding to a 90\% confidence level and a 10\% margin of error, and were independently evaluated by two of the authors.

\textbf{Expert study:} To reduce potential author bias and to obtain a practitioner perspective, a random subset of 14 of these 64 instances, corresponding to a 90\% confidence level and a 20\% margin of error, was additionally evaluated by external experts: a senior software architect with more than 10 years of industry experience, and two PhD students with 2 years of industry experience, and researching software architecture (hereafter referred to as junior architects). The senior architect evaluated all 14 instances, while each junior architect evaluated a disjoint half (7 instances each). Since the two halves do not overlap, we treat the two PhDs as a single \emph{junior architect} rater, so that every instance in the expert study received exactly one senior and one junior rating.

\textbf{Model Selection:} 
The evaluation dataset was constructed by sampling 64 keys from the test set. For each approach, the top-performing model was selected based on automated BERTScore evaluation on the test set. The selected models were:
\begin{itemize}
    \item Prompting: Gemini 2.5 Flash for both CD and TB tasks
    \item Finetuning: Qwen3 30B A3B Instruct for both CD and TB tasks.
    \item RAFG: Qwen3 for the CD task, and Gemma 3 4B for the TB task.
    \item DRAFT: Qwen3 30B A3B Instruct for both CD and TB tasks.
\end{itemize}

\textbf{Evaluation Platform:}
To facilitate the study, two custom Google Web Apps were built corresponding to the different tasks (CD and TB). Each application presented the evaluators with the source material, which consisted of the Context and ground truth Decision for CD tasks, or the Title and ground truth Body for TB tasks. Below this source material, the apps presented four generated outputs—one from each approach—which were blinded and randomly shuffled to ensure an unbiased assessment. Authors and experts used the same applications and were shown the same information, including the ground truth, which is necessary to judge closeness to the reference.

\textbf{Evaluation Metrics:}
For each generated output, human evaluators assigned two separate 5-star ratings based on the following criteria:
\begin{itemize}
    \item Architectural Correctness: An assessment of how well the generated decision or ADR follows sound software architecture principles.
    \item Closeness to Ground Truth: An assessment of how similar the generated output is to the reference solution.
\end{itemize}
  

Following data collection, the human ratings were aggregated to compute mean scores for each approach, inter-rater agreement, and statistical significance tests. These were computed separately for the author study, the expert study, and all four raters on the 14 instances shared by both studies (Table \ref{tab:humanEval}).


\medskip
\noindent \textbf{Statistical Significance tests}

We assess the statistical significance of the results using paired per-sample metric scores. For each task–model pair, we first apply the Friedman test~\cite{friedman1937} to determine whether the experimental approaches differ significantly overall. When the omnibus test is significant, we perform pairwise Wilcoxon signed-rank tests~\cite{wilcoxon1945} and adjust the resulting p-values using the Holm correction~\cite{holm1979}. We report Kendall’s W~\cite{kendall1939} for the Friedman test and rank-biserial correlation for pairwise comparisons~\cite{kerby2014} as effect-size measures. This analysis is conducted for all automatic evaluation metrics, including BLEU, ROUGE-1, ROUGE-L, METEOR, and BERTScore (precision, recall, and F1).


To rigorously assess the human evaluation scores, we employed non-parametric statistical methods appropriate for ordinal Likert scale data. First, we calculated inter rater reliability using quadratic Weighted Cohen's Kappa ($\kappa$)~\cite{cohen1968}. This metric was chosen over standard agreement measures because it applies quadratic weights to penalize severe evaluator disagreements, which is particularly useful for subjective generation metrics. For the author study, $\kappa$ is computed between the two authors, and for the expert study, between the senior and the junior architect. When all four raters are considered together, we report the mean of the six pairwise $\kappa$ values. For performance comparisons, the ratings of all raters who evaluated an instance were averaged per instance, and we used the Friedman test as an omnibus measure to detect statistically significant differences among the four approaches (Prompting, \RAG, \Finetuning, and \Approach) on these matched samples ($n=64$ for the author study, $n=14$ for the expert study and for all raters combined). When the Friedman test indicated a significant difference ($p < 0.05$), we conducted pairwise Wilcoxon signed-rank tests comparing \Approach~against each baseline, with Holm correction.

\end{rebuttalblue}


\subsection{RQ1 : Effectiveness of Traditional Methods} \label{subsec:rq1}
To assess the effectiveness of \Approach, we first evaluate how well other popular approaches can generate Architectural artifacts that align with real-world architectural documentation as described in Section \ref{subsec:expCandidates}. Similarity is measured by comparing the generated outputs against ground truth \Decision s and ADR Bodies from actual repositories.

\medskip
\noindent \textbf{Prompting}

We evaluated all the models on the test set for Prompting, which forms our baseline for both tasks.

\textbf{Task 1 (Context → Decision):} As reported in Table \ref{tab:results_cd}, Gemini-2.5 achieved strong performance in prompting, with a ROUGE-1 score of 0.162, BLEU score of 0.015, METEOR score of 0.158, and BERTScore F1 of 0.829. Qwen-3-30B also performed well with a BERTScore F1 of 0.822. These results indicate reasonable alignment with the reference Design Decisions. Conversely, Gemma-3-4B scored the lowest among the models tested with a BERTScore F1 of 0.805.

\textbf{Task 2 (Title → Body):} As shown in Table \ref{tab:results_tb}, Qwen-3-30B demonstrated strong baseline performance with a ROUGE-1 score of 0.242, BLEU score of 0.026, METEOR score of 0.226, and BERTScore F1 of 0.817. Gemini-2.5 achieved a comparable BERTScore F1 of 0.815. For this task, Gpt-5 had the lowest performance with a BERTScore F1 of 0.803, though this was still reasonable given the more complex nature of generating full ADR bodies.

\medskip
\noindent \textbf{\RAFewshotG}

Results for both tasks, summarized in Tables \ref{tab:results_cd} and \ref{tab:results_tb} and graphically depicted in Figures \ref{fig:CD_Results} and \ref{fig:TB_Results}, reveal that \RAG~improves performance for most LLMs across both tasks, though the magnitude of improvement varies by model and task.

\textbf{Task 1 (Context → Decision):} \RAG~had a clear positive impact on the performance of most models. Gemma-3-4b showed the most substantial improvement with BERTScore F1 increasing from 0.805 to 0.843. The proprietary models also benefited. GPT-5 showed improvement with BERTScore F1 increasing from 0.812 to 0.815, and Gemini-2.5 improved from 0.829 to 0.834. Qwen3-30B showed consistent improvement across metrics, with BERTScore F1 rising from 0.822 to 0.847. These results indicate that retrieval-augmented examples help models generate Architectural artifacts that better align with real repository documentation.

\textbf{Task 2 (Title → Body):} For the more complex task of generating complete ADR Bodies, \RAG~continued to show benefits, though with more varied patterns. Gemma-3-4b and GPT-5 demonstrated improvements, with Gemma-3-4b's BERTScore F1 rising from 0.814 to 0.822. Qwen3-30B maintained its baseline performance (BERTScore F1 of 0.817), while Gemini-2.5 showed a slight decrease in BERTScore F1 from 0.815 to 0.813, though it improved on ROUGE-1. This suggests that for some models, retrieval-augmented examples for longer, more complex outputs may occasionally introduce conflicting patterns.

\medskip
Our observations indicate that \RAG~improves the similarity of LLM-generated outputs to real repository documentation across model architectures. The improvement is more consistent and pronounced for the simpler Task 1, with notable gains for Gemma-3-4b. For the more complex Task 2, performance gains are smaller but still observable.

\medskip
\noindent \textbf{\Finetuning}

As seen in Tables \ref{tab:results_cd} and \ref{tab:results_tb}, and Figures \ref{fig:CD_Results} and \ref{fig:TB_Results}, \finetuning~had a generally positive impact on model performance across both tasks, with most models demonstrating substantial improvement in BERTScore F1 after \finetuning.

\textbf{Task 1 (Context → Decision):} Fine-tuning showed strong improvements across all open-source models. Gemma-3-4b showed notable increase in BERTScore F1 from 0.805 to 0.834. Qwen3-30B achieved a BERTScore F1 of 0.847, up from a baseline of 0.822. The improvements were consistent across other metrics as well, with ROUGE-1 scores increasing from 0.135 to 0.186 for Gemma-3-4b and from 0.159 to 0.210 for Qwen3-30B. Interestingly, Gemini-2.5 showed a slight decrease in BERTScore F1 from 0.829 to 0.827, though this difference is marginal.

\textbf{Task 2 (Title → Body):} \Finetuning~showed even more pronounced improvements for Task 2, which is a more complex generation task. Gemma-3-4b achieved a BERTScore F1 of 0.830, up from 0.814. Qwen3-30B showed substantial improvement with BERTScore F1 rising from 0.817 to 0.835. Gemini-2.5 also showed improvement from 0.815 to 0.820. The ROUGE-1 scores showed significant gains as well, with Gemma-3-4b improving from 0.221 to 0.269 and Gemini-2.5 from 0.145 to 0.182.

\begin{table}[t]
\centering
\small
\color{black}

\resizebox{\linewidth}{!}{
\begin{tabular}{llcccccc}
\toprule
Metric & Model & Friedman & \Approach-Finetune & \Approach-Prompting & \Approach-RAFG & Finetune-Prompting & RAFG-Prompting \\
\midrule

\multirow{3}{*}{ROUGE-1} & Gemini-2.5 & 0.087 *** & 0.533 *** & 0.266 *** & 0.147 *** & -0.398 *** & 0.167 *** \\
 & Gemma-3-4b & 0.088 *** & -0.061 ns & 0.461 *** & -0.100 * & 0.476 *** & 0.479 *** \\
 & Qwen-3-30b & 0.033 *** & -0.011 ns & 0.247 *** & 0.082 ns & 0.274 *** & 0.525 *** \\
\midrule

\multirow{3}{*}{ROUGE-L} & Gemini-2.5 & 0.040 *** & 0.353 *** & 0.307 *** & 0.174 *** & -0.163 *** & 0.184 *** \\
 & Gemma-3-4b & 0.142 *** & -0.077 ns & 0.599 *** & -0.166 *** & 0.622 *** & 0.682 *** \\
 & Qwen-3-30b & 0.096 *** & -0.013 ns & 0.551 *** & 0.399 *** & 0.565 *** & 0.591 *** \\
\midrule

\multirow{3}{*}{BLEU} & Gemini-2.5 & 0.005 ** & 0.404 *** & 0.483 *** & 0.293 *** & 0.017 ns & 0.254 ** \\
 & Gemma-3-4b & 0.003 ns & 0.026 ns & 0.387 *** & 0.313 *** & 0.404 *** & 0.137 ns \\
 & Qwen-3-30b & 0.002 ns & 0.028 ns & 0.358 *** & 0.371 *** & 0.323 *** & 0.116 ns \\
\midrule

\multirow{3}{*}{METEOR} & Gemini-2.5 & 0.161 *** & 0.604 *** & 0.093 * & -0.038 ns & -0.579 *** & 0.191 *** \\
 & Gemma-3-4b & 0.008 *** & -0.023 ns & -0.127 ** & 0.038 ns & -0.114 * & -0.165 *** \\
 & Qwen-3-30b & 0.127 *** & -0.034 ns & -0.320 *** & -0.389 *** & -0.336 *** & 0.056 ns \\
\midrule

\multirow{3}{*}{BERT-Precision} & Gemini-2.5 & 0.051 *** & 0.023 ns & 0.341 *** & 0.209 *** & 0.333 *** & 0.266 *** \\
 & Gemma-3-4b & 0.513 *** & -0.129 ** & 0.896 *** & -0.526 *** & 0.918 *** & 0.999 *** \\
 & Qwen-3-30b & 0.579 *** & 0.080 * & 0.942 *** & 0.873 *** & 0.894 *** & 0.883 *** \\
\midrule

\multirow{3}{*}{BERT-Recall} & Gemini-2.5 & 0.052 *** & 0.451 *** & 0.202 *** & 0.073 ns & -0.306 *** & 0.163 *** \\
 & Gemma-3-4b & 0.030 *** & -0.052 ns & 0.202 *** & -0.153 *** & 0.245 *** & 0.323 *** \\
 & Qwen-3-30b & 0.022 *** & -0.023 ns & -0.099 * & -0.225 *** & -0.074 ns & 0.212 *** \\
\midrule

\multirow{3}{*}{BERT-F1} & Gemini-2.5 & 0.043 *** & 0.315 *** & 0.380 *** & 0.215 *** & -0.062 ns & 0.271 *** \\
 & Gemma-3-4b & 0.402 *** & -0.108 ** & 0.828 *** & -0.467 *** & 0.847 *** & 0.972 *** \\
 & Qwen-3-30b & 0.340 *** & 0.055 ns & 0.825 *** & 0.685 *** & 0.795 *** & 0.750 *** \\
\bottomrule
\end{tabular}}

\caption{\rebuttal{Statistical significance of approach comparisons for Task 1. We report the effect sizes along with significance levels for Friedman omnibus test followed by Wilcoxon signed-rank tests with Holm correction. Significance levels: *** ($p<0.001$), ** ($p<0.01$), * ($p<0.05$), ns = not significant.}}
\label{tab:significance_task1}

\end{table}

\begin{rebuttalblue}

\medskip
\noindent \textbf{Human Evaluations}\\

The author ratings (Table \ref{tab:humanEval}) corroborate the automated metrics for the baseline approaches, with qualitative feedback highlighting specific generational flaws. For both Task 1 and Task 2, \RAG~and \Finetuning~performed better than zero-shot Prompting across both evaluation dimensions. In Task 1, Prompting yielded the lowest author ratings (Closeness: 2.89, Correctness: 3.38), whereas both \RAG~and \Finetuning~improved upon these baseline scores (\RAG~Correctness: 3.55; \Finetuning~Correctness: 3.48). Evaluator comments revealed that Prompting frequently suffered from extreme verbosity and structural failure, describing outputs as an ``insanely long decision for a simple ADR entry'' or simply ``Generation incomplete.'' In the more complex Task 2, \Finetuning~emerged as the stronger baseline (Closeness: 2.63, Correctness: 3.09), but evaluators noted a strong tendency for the model to hallucinate extraneous details, remarking that the ``Architecture is correct, but then it starts hallucinating''. \RAG~mitigated some completeness issues but occasionally generated outputs ``Needlessly repeating the context''.

The expert ratings present a different ordering among the baselines. Experts rated \RAG~the highest of all approaches on both tasks, particularly for Correctness (3.82 for Task 1 and 3.57 for Task 2), while \Finetuning~received lower ratings than Prompting (e.g., Task 2 Correctness: 2.32 vs. 2.50). The experts' comments echo the authors' observations on Prompting, describing outputs that ``stop midway'' or are ``not complete''. \RAG~outputs were described as ``more elaborate'' and as explaining ``how the data problem can be addressed'', although they sometimes ``invent unsupported details''. For Task 2, the junior architects repeatedly noted that ``the title alone is not sufficient to derive'' the specifics of the ground truth, which is consistent with the low Closeness scores across all approaches for this task.

\medskip
\noindent \textbf{Statistical significance}

To determine whether the improvements from Retrieval Augmented Few-shot Generation and fine-tuning reported above are statistically robust rather than attributable to per-sample variance, we apply the procedure described in Section~\ref{subsec:Evaluation} to the paired, per-sample metric scores. For nearly every metric--model combination in both tasks, the Friedman omnibus test rejects the null hypothesis that all four approaches perform equivalently ($p < 0.001$; Tables~\ref{tab:significance_task1} and~\ref{tab:significance_task2}), with Kendall's $W$ ranging from as low as $W = 0.002$ up to $W = 0.579$. The main exception is BLEU on Task~1, where the omnibus test is not significant for Gemma-3-4b ($W = 0.003$, n.s.) or Qwen3-30B ($W = 0.002$, n.s.), indicating that approach choice does not meaningfully shift BLEU scores for these two models, consistent with BLEU's known sensitivity to short, low-$n$-gram-overlap outputs such as single-sentence \Decision.

\textbf{Task~1 (Context $\rightarrow$ Decision):} Fine-tuning produces large, highly significant gains over prompting for the open-source models: BERT-Precision improves with near-maximal effect sizes for Gemma-3-4b ($r = 0.918$, $p < 0.001$) and Qwen3-30B ($r = 0.894$, $p < 0.001$), with correspondingly large BERT-F1 effects ($r = 0.847$ and $r = 0.795$, respectively). For Gemini-2.5, however, fine-tuning is significantly \emph{worse} than prompting on ROUGE-1 ($r = -0.398$, $p < 0.001$), METEOR ($r = -0.579$, $p < 0.001$), and BERT-Recall ($r = -0.306$, $p < 0.001$), while the BERT-F1 difference itself is not significant ($r = -0.062$, n.s.) confirming that the marginal BERT-F1 decrease reported descriptively for Gemini-2.5 is not a reliable effect, while revealing that fine-tuning does measurably hurt Gemini-2.5's lexical alignment, a pattern the BERTScore comparison alone would miss. RAFG shows the most consistent improvement over prompting of the three baselines: it outperforms prompting with near-maximal effect sizes on BERT-Precision for Gemma-3-4b ($r = 0.999$, $p < 0.001$) and Qwen3-30B ($r = 0.883$, $p < 0.001$), and large effects on BERT-F1 for both ($r = 0.972$ and $r = 0.750$, respectively).

\textbf{Task~2 (Title $\rightarrow$ Body):} Fine-tuning is significantly better than prompting on BERT-F1 for all three models, with large effects for Gemma-3-4b ($r = 0.729$, $p < 0.001$) and Qwen3-30B ($r = 0.647$, $p < 0.001$) and a smaller but still significant effect for Gemini-2.5 ($r = 0.262$, $p < 0.001$), corroborating the stronger Task~2 gains. RAFG's advantage is less consistent here: it remains strong for Gemma-3-4b (BERT-F1 $r = 0.761$, $p < 0.001$), effectively disappears for Qwen3-30B ($r = 0.016$, n.s.), and reverses for Gemini-2.5, which is significantly worse under RAFG than prompting ($r = -0.092$, $p < 0.05$). This confirms that RAFG's benefit on Task~2, is not uniform once model and task are held constant. 
\end{rebuttalblue}

\medskip
These results demonstrate that \finetuning~consistently enhances the ability of LLMs to generate outputs that closely align with real repository documentation across both tasks, with particularly strong improvements for the more complex Task 2. This validates the positive impact of domain-specific \finetuning~of LLMs in supporting architectural decision making.
\subsection{RQ2 : Effectiveness of \Approach~} \label{subsec:rq2}

In RQ1 we found that both \RAFewshotG~and \finetuning~usually increases the similarity of LLM generated Architectural artifacts to real repository documentation. Building upon this insight we designed \Approach~as described in Section \ref{sec:Overview} and \ref{sec:Approach}.

The automated metrics, summarized in Tables \ref{tab:results_cd} and \ref{tab:results_tb}, and Figures \ref{fig:CD_Results} and \ref{fig:TB_Results}, demonstrate that \Approach~significantly improves the alignment of LLM-generated outputs with real repository documentation across both tasks.

\textbf{Task 1 (Context → Decision):} As shown in Table \ref{tab:results_cd} and Figure \ref{fig:CD_Results}, \Approach~achieved the best overall results. Qwen3-30B with \Approach~achieved a ROUGE-1 score of 0.213, BLEU score of 0.041, METEOR score of 0.167, and BERTScore F1 of 0.850, representing the highest performance across all approaches for this model. Gemini-2.5 with \Approach~also showed strong performance with a BERTScore F1 of 0.839, an improvement over both \finetuning~(0.827) and \RAG~(0.834) performance. Gemma-3-4b achieved a BERTScore F1 of 0.831 with \Approach, which while strong, was slightly lower than its \RAG~performance (0.843), suggesting that for some smaller models, the benefits of combining both techniques may vary.

\textbf{Task 2 (Title → Body):} As shown in Table \ref{tab:results_tb} and Figure \ref{fig:TB_Results}, \Approach~demonstrated even more substantial improvements for the more complex task of generating ADR bodies. Qwen3-30B with \Approach~achieved a ROUGE-1 score of 0.251, BLEU score of 0.035, METEOR score of 0.194, and BERTScore F1 of 0.839, representing the best overall performance. Gemma-3-4b with \Approach~showed improvement with a BERTScore F1 of 0.832, up from its prompting baseline of 0.814. Most notably, Gemini-2.5 with \Approach~achieved a BERTScore F1 of 0.830, a significant improvement over \finetuning~(0.820) and \RAG~(0.813) performance.
A noteworthy observation from Figure \ref{fig:TB_Results} is that while \Approach~consistently leads in BERTScore F1, \finetuning~and \RAG~occasionally achieve the highest results in ROUGE-1, BLEU, and METEOR for specific model-task combinations. Notably, the peak scores for these lexical metrics are often reached through the \finetuning~of~\RAG.

\medskip
\noindent \textbf{Cross-Task Analysis:}\\
Comparing performance across tasks, we observe that the absolute metric scores are generally higher for Task 2, reflecting the longer and more information rich nature of ADR bodies compared to decisions. Additionally, \Approach~'s improvements over baseline are more pronounced for Task 2, suggesting that the combination of \RAG~and \finetuning~is particularly beneficial for more complex generation tasks.

The relative ranking of approaches is consistent across both tasks, with \Approach~and \finetuning~generally outperforming prompting and \RAG~alone. This finding suggests that \finetuning~or \Approach -ing effectively adapts the model to domain specific settings, particularly in the case of AKM.

Overall, these results demonstrate that \Approach -ing increases the effectiveness of LLMs in generating ADRs that align more closely with real repository documentation compared to prompting, \finetuning, and \RAG~used individually, across both tasks. Integrating \finetuning~with \RAFewshotG~substantially improves the LLM’s ability to capture the structure, style, and content of human authored ADRs. These gains are especially pronounced for the more complex Task 2.

\begin{table}[ht]
\centering
\small
\color{blue}

\resizebox{\linewidth}{!}{
\begin{tabular}{llcccccc}
\toprule
Metric & Model & Friedman & \Approach-Finetune & \Approach-Prompting & \Approach-RAFG & Finetune-Prompting & RAFG-Prompting \\
\midrule

\multirow{3}{*}{ROUGE-1} & Gemini-2.5 & 0.084 *** & 0.275 *** & 0.552 *** & 0.487 *** & 0.327 *** & 0.124 ** \\
 & Gemma-3-4b & 0.141 *** & -0.063 ns & 0.446 *** & 0.197 *** & 0.594 *** & 0.680 *** \\
 & Qwen-3-30b & 0.004 * & 0.109 * & 0.041 ns & 0.042 ns & -0.080 ns & -0.018 ns \\
\midrule

\multirow{3}{*}{ROUGE-L} & Gemini-2.5 & 0.101 *** & 0.287 *** & 0.561 *** & 0.511 *** & 0.321 *** & 0.091 * \\
 & Gemma-3-4b & 0.290 *** & -0.024 ns & 0.739 *** & 0.480 *** & 0.842 *** & 0.803 *** \\
 & Qwen-3-30b & 0.140 *** & 0.067 ns & 0.666 *** & 0.668 *** & 0.560 *** & -0.000 ns \\
\midrule

\multirow{3}{*}{BLEU} & Gemini-2.5 & 0.109 *** & 0.156 *** & 0.682 *** & 0.324 *** & 0.586 *** & 0.515 *** \\
 & Gemma-3-4b & 0.306 *** & -0.074 ns & 0.821 *** & 0.298 *** & 0.886 *** & 0.941 *** \\
 & Qwen-3-30b & 0.080 *** & -0.053 ns & 0.498 *** & 0.484 *** & 0.532 *** & 0.099 ns \\
\midrule

\multirow{3}{*}{METEOR} & Gemini-2.5 & 0.073 *** & 0.147 *** & 0.547 *** & 0.229 *** & 0.445 *** & 0.390 *** \\
 & Gemma-3-4b & 0.098 *** & -0.359 *** & -0.079 * & -0.445 *** & 0.238 *** & 0.686 *** \\
 & Qwen-3-30b & 0.143 *** & -0.030 ns & -0.441 *** & -0.438 *** & -0.459 *** & 0.009 ns \\
\midrule

\multirow{3}{*}{BERT-Precision} & Gemini-2.5 & 0.102 *** & 0.386 *** & 0.341 *** & 0.582 *** & -0.100 * & -0.373 *** \\
 & Gemma-3-4b & 0.329 *** & 0.215 *** & 0.862 *** & 0.721 *** & 0.771 *** & 0.647 *** \\
 & Qwen-3-30b & 0.471 *** & 0.173 *** & 0.930 *** & 0.929 *** & 0.833 *** & 0.038 ns \\
\midrule

\multirow{3}{*}{BERT-Recall} & Gemini-2.5 & 0.083 *** & 0.190 *** & 0.563 *** & 0.323 *** & 0.449 *** & 0.221 *** \\
 & Gemma-3-4b & 0.124 *** & -0.094 * & 0.351 *** & -0.202 *** & 0.468 *** & 0.802 *** \\
 & Qwen-3-30b & 0.037 *** & -0.050 ns & -0.221 *** & -0.216 *** & -0.198 *** & -0.005 ns \\
\midrule

\multirow{3}{*}{BERT-F1} & Gemini-2.5 & 0.135 *** & 0.450 *** & 0.658 *** & 0.654 *** & 0.262 *** & -0.092 * \\
 & Gemma-3-4b & 0.252 *** & 0.139 *** & 0.780 *** & 0.490 *** & 0.729 *** & 0.761 *** \\
 & Qwen-3-30b & 0.261 *** & 0.135 ** & 0.792 *** & 0.785 *** & 0.647 *** & 0.016 ns \\
\bottomrule
\end{tabular}}

\caption{\rebuttal{Statistical significance of approach comparisons for Task 2. We report the effect sizes along with significance levels for Friedman omnibus test followed by Wilcoxon signed-rank tests with Holm correction. Significance levels: *** ($p<0.001$), ** ($p<0.01$), * ($p<0.05$), ns = not significant.}}
\label{tab:significance_task2}

\end{table}

\begin{rebuttalblue}


\medskip
\noindent \textbf{Human Evaluations}\\
The human evaluation results (Table \ref{tab:humanEval}) differ between rater groups. In the author study, \Approach~received the highest mean ratings on both tasks and both dimensions (Task 1 Closeness: 3.20, Correctness: 3.68; Task 2 Closeness: 2.72, Correctness: 3.25). Authors noted that \Approach~generated the most structurally sound outputs, although they occasionally flagged internal contradictions, noting instances where a generation ``conflicts with itself'' or ``Hallucinated a bit.''

In the expert study, however, \RAG~received the highest ratings, and \Approach~was rated below the baselines on Task 1 (Closeness: 2.68, Correctness: 2.82). When all four raters are combined on the 14 shared instances, no approach is consistently preferred: \RAG~leads on Task 1 and on Task 2 Correctness, \Approach~leads on Task 2 Closeness (2.57), and the differences between \Approach, \Finetuning, and Prompting are small.

A rater-level analysis indicates that the expert result is driven largely by the senior architect. The senior architect rated \Approach~low on Task 1 (Closeness: 2.07, Correctness: 2.36) and strongly preferred \RAG~on Correctness in both tasks (Task 1: 4.21; Task 2: 4.14 vs. at most 2.14 for every other approach). In contrast, the junior architects found little difference between the approaches: their mean ratings lie within 0.6 points of each other in every setting, and they gave \Approach~the highest Closeness rating on Task 1 (3.29). Their comments also credit specific strengths of \Approach. Several outputs were described as capturing the core decision, for example ``This is same as ground truth'', and ``captures the main idea of moving expensive git operations to a background worker''. The most frequent criticism was brevity: \Approach~outputs were described as ``very concise'' but missing rationale (``the rationale is not mentioned''), or as ``a blanket statement''. This is consistent with Section \ref{subsec:rq3}, where \Approach~produces the shortest outputs (e.g., 61 tokens for Qwen3-30B on Task 1).


Across all rater groups and approaches, Architectural Correctness was rated higher than Closeness, indicating that LLMs frequently generate architecturally valid alternatives even when they deviate from the ground truth.

\medskip
\noindent \textbf{Statistical Significance of Automated Evaluation}\\
We apply the same paired significance procedure to test whether \Approach's improvements over each baseline are statistically reliable rather than apparent only in the mean. In the pairwise tests of \Approach-Finetune, \Approach-Prompting, and \Approach-RAFG columns of Tables~\ref{tab:significance_task1} and~\ref{tab:significance_task2}, a positive rank-biserial correlation $r$ indicates \Approach~outperforms the comparison approach, and a negative value indicates the reverse.

\textbf{Task~1 (Context $\rightarrow$ Decision):} \Approach~significantly outperforms prompting on BERT-F1 for every model, with large effect sizes throughout Gemma-3-4b ($r = 0.828$, $p < 0.001$), Qwen3-30B ($r = 0.825$, $p < 0.001$), and Gemini-2.5 ($r = 0.380$, $p < 0.001$), confirming that \Approach's advantage over the weakest baseline is both consistent and substantial. The comparison against fine-tuning and RAFG alone is more nuanced and model-dependent. For Gemini-2.5, \Approach~shows a significant, large improvement over both fine-tuning (BERT-F1 $r = 0.315$, $p < 0.001$) and RAFG ($r = 0.215$, $p < 0.001$). For Qwen3-30B, \Approach~is statistically indistinguishable from fine-tuning alone on BERT-F1 ($r = 0.055$, n.s.) but significantly outperforms RAFG alone with a large effect ($r = 0.685$, $p < 0.001$). For Gemma-3-4b, the pattern reverses: \Approach~is significantly \emph{outperformed} by both fine-tuning ($r = -0.108$, $p < 0.01$) and RAFG ($r = -0.467$, $p < 0.001$) on BERT-F1. This provides direct statistical support for the qualitative observation in Section~\ref{subsec:LessonsLearned} that \Approach's gains are least reliable for the smallest model in the study.

\textbf{Task~2 (Title $\rightarrow$ Body):} The picture shifts substantially. \Approach~again significantly outperforms prompting for all three models on BERT-F1, with large effects (Gemma-3-4b $r = 0.780$; Qwen3-30B $r = 0.792$; Gemini-2.5 $r = 0.658$; all $p < 0.001$). Critically, \Approach~now also significantly outperforms both remaining baselines for every model: against fine-tuning, Gemini-2.5 ($r = 0.450$, $p < 0.001$), Gemma-3-4b ($r = 0.139$, $p < 0.001$), and Qwen3-30B ($r = 0.135$, $p < 0.01$); against RAFG, Gemini-2.5 ($r = 0.654$, $p < 0.001$), Gemma-3-4b ($r = 0.490$, $p < 0.001$), and Qwen3-30B ($r = 0.785$, $p < 0.001$). Notably, Gemma-3-4b's relationship to \finetuning~inverts between tasks, significantly favoring \finetuning on Task~1 ($r = -0.108$) but significantly favoring \Approach~on Task~2 ($r = 0.139$) offering statistical confirmation, for the first time at the level of individual pairwise tests, that \Approach's benefit over its constituent techniques is concentrated in the more complex, longer-form Task~2 rather than being a uniform property of the approach itself.
\end{rebuttalblue}




\begin{table*}[ht]
\centering
\color{blue}
\setlength{\tabcolsep}{5pt}
\begin{tabular*}{\textwidth}{@{\extracolsep{\fill}}llcccc}
\toprule
\multirow{2}{*}{\textbf{Raters}} & \multirow{2}{*}{\textbf{Approach}} & \multicolumn{2}{c}{\textbf{Task 1}} & \multicolumn{2}{c}{\textbf{Task 2}} \\
\cmidrule(lr){3-4} \cmidrule(lr){5-6}
 & & \textbf{Closeness} & \textbf{Correctness} & \textbf{Closeness} & \textbf{Correctness} \\
\midrule
\multirow{6}{*}{\shortstack[l]{Authors\\($n=64$)}}
 & Prompting  & 2.89 & 3.38\rlap{$^{*}$} & 2.41 & 2.98 \\
 & RAFG       & 2.99 & 3.55 & 2.44 & 3.09 \\
 & Finetuning & 2.95 & 3.48 & 2.63 & 3.09 \\
 & DRAFT      & \textbf{3.20} & \textbf{3.68} & \textbf{2.72} & \textbf{3.25} \\
\cmidrule(lr){2-6}
 & Friedman $p$       & 0.049 & 0.028 & 0.065 & 0.165 \\
 & $\kappa$           & 0.352 & 0.205 & 0.258 & 0.259 \\
\midrule
\multirow{6}{*}{\shortstack[l]{Experts\\($n=14$)}}
 & Prompting  & 2.96 & 3.46 & \textbf{2.46} & 2.50 \\
 & RAFG       & \textbf{3.39} & \textbf{3.82} & \textbf{2.46} & \textbf{3.57}\rlap{$^{*}$} \\
 & Finetuning & 2.75 & 2.93 & 2.11 & 2.32 \\
 & DRAFT      & 2.68 & 2.82 & 2.29 & 2.32 \\
\cmidrule(lr){2-6}
 & Friedman $p$       & 0.169 & 0.031 & 0.239 & 0.002 \\
 & $\kappa$           & 0.569 & 0.472 & 0.683 & 0.407 \\
\midrule
\multirow{6}{*}{\shortstack[l]{All raters\\($n=14$)}}
 & Prompting  & 2.89 & 3.36 & 2.45 & 2.89 \\
 & RAFG       & \textbf{3.29} & \textbf{3.64} & 2.55 & \textbf{3.39} \\
 & Finetuning & 2.89 & 3.23 & 2.32 & 2.82 \\
 & DRAFT      & 2.96 & 3.30 & \textbf{2.57} & 2.84 \\
\cmidrule(lr){2-6}
 & Friedman $p$       & 0.196 & 0.151 & 0.601 & 0.421 \\
 & ${\kappa}^{\dagger}$ & 0.368 & 0.267 & 0.374 & 0.233 \\
\bottomrule
\end{tabular*}
\caption{\rebuttal{Human evaluation mean scores (1--5) for Task 1 and Task 2. \emph{Authors}: two authors rated 64 samples. \emph{Experts}: a senior architect and junior architects rated a 14-sample subset. \emph{All raters}: all four raters on the same 14 samples. Bold marks the highest mean within each rater group. $^{*}$ indicates a statistically significant difference from DRAFT (Wilcoxon signed-rank, Holm-corrected, $p<0.05$), tested only when the Friedman test is significant. $\kappa$: quadratic weighted Cohen's kappa between the two raters. $^{\dagger}$Mean of the six pairwise weighted kappas among the four raters.}}
\label{tab:humanEval}
\end{table*}

\begin{rebuttalblue}

\medskip
\noindent \textbf{Statistical Significance of Human Evaluation}\\
Inter-rater agreement between the two authors ranged from $\kappa=0.205$ to $0.352$, denoting a fair level of agreement, which is common for subjective architectural assessments. Agreement between the senior and the junior architect was higher, ranging from $\kappa=0.407$ to $0.683$ (moderate to substantial). Agreement between the author and expert groups was lower (mean pairwise $\bar{\kappa}$ between 0.233 and 0.374 across the four raters), underlining that judgements of architectural quality depend on the evaluator.

In the author study ($n=64$), the Friedman test detected significant differences for both Closeness ($p=0.049$) and Correctness ($p=0.028$) in Task 1. Subsequent pairwise Wilcoxon signed-rank tests with Holm-Bonferroni correction showed that \Approach~significantly outperforms zero-shot Prompting in Correctness ($p=0.024$). In Task 2, although \Approach~led in absolute performance, the Friedman tests did not reach significance ($p=0.065$ for Closeness; $p=0.165$ for Correctness). This may be because evaluators sometimes assigned equal ratings to all approaches, noting that ``the original title is actually too small or generic to mean anything.''

In the expert study ($n=14$), the Friedman test was significant for Correctness in both tasks (Task 1: $p=0.031$; Task 2: $p=0.002$) but not for Closeness. In Task 1, no pairwise comparison against \Approach~remained significant after correction. In Task 2, \RAG~was rated significantly higher than \Approach~in Correctness ($p=0.011$). Analysed per rater, these significant differences stem from the senior architect (Friedman $p=0.010$ and $p<0.001$ for Correctness in Task 1 and Task 2), whereas none of the junior architects' ratings showed a significant difference between approaches (all $p>0.25$). When the ratings of all four raters are combined, no significant differences remain in either task (Friedman $p$ between 0.151 and 0.601).


\end{rebuttalblue}




\subsection{\textbf{RQ3} Efficiency of \Approach} \label{subsec:rq3}

\begin{table*}[ht]
\centering
\scriptsize
\setlength{\tabcolsep}{4pt}

\begin{tabular}{
>{\centering\arraybackslash}p{0.11\textwidth}
>{\centering\arraybackslash}p{0.11\textwidth}
|>{\centering\arraybackslash}p{0.07\textwidth}
>{\centering\arraybackslash}p{0.07\textwidth}
>{\centering\arraybackslash}p{0.08\textwidth}
|>{\centering\arraybackslash}p{0.07\textwidth}
>{\centering\arraybackslash}p{0.07\textwidth}
>{\centering\arraybackslash}p{0.08\textwidth}
}
\toprule
\multirow{2}{*}{\textbf{Model}} &
\multirow{2}{*}{\textbf{Approach}} &
\multicolumn{3}{c|}{\textbf{(a) Task 1: C $\rightarrow$ D}} &
\multicolumn{3}{c}{\textbf{(b) Task 2: T $\rightarrow$ B}} \\
\cmidrule(lr){3-5}\cmidrule(lr){6-8}
& &
\textbf{Input} & \textbf{Output} & \textbf{Time (s)} &
\textbf{Input} & \textbf{Output} & \textbf{Time (s)} \\
\midrule

\multirow{4}{*}{Gemma-3-4b}
& Prompting  & 122 & 395 & \cellcolor{worst}26.08 & 8   & 981 & \cellcolor{worst}32.42 \\
& Finetuning & 122 & 128 & 4.29                     & 8   & 453 & 15.14 \\
& RAFG       & 559 & 79  & \cellcolor{best}3.88    & 811 & 680 & 25.48 \\
& \Approach  & 559 & 140 & 5.17                     & 811 & 313 & \cellcolor{best}11.51 \\
\midrule

\multirow{4}{*}{Qwen3-30B}
& Prompting  & 122 & 275 & \cellcolor{worst}74.82  & 8   & 500 & 155.20 \\
& Finetuning & 122 & 87  & 39.46                    & 8   & 264 & 110.14 \\
& RAFG       & 559 & 171 & 59.93                    & 811 & 500 & \cellcolor{worst}177.11 \\
& \Approach  & 559 & 61  & \cellcolor{best}28.97   & 811 & 182 & \cellcolor{best}75.51 \\
\midrule

\multirow{4}{*}{gemini-2.5}
& Prompting  & 122 & 147 & \cellcolor{worst}6.27   & 8   & 155 & \cellcolor{worst}7.55 \\
& Finetuning & 122 & 122 & \cellcolor{best}1.31    & 8   & 306 & \cellcolor{best}6.13 \\
& RAFG       & 527 & 151 & 6.03                     & 759 & 312 & 7.33 \\
& \Approach  & 527 & 135 & 1.73                     & 759 & 264 & 6.92 \\
\midrule

\multirow{2}{*}{gpt-5}
& Prompting  & 122 & 672 & 17.19                    & 8   & 727 & 17.25 \\
& RAFG       & 643 & 605 & 14.74                    & 801 & 695 & 15.44 \\
\bottomrule
\end{tabular}

\caption{Efficiency results reporting mean input tokens, output tokens, and response time per model--approach combination for (a) Task 1 (C $\rightarrow$ D) and (b) Task 2 (T $\rightarrow$ B).}
\label{tab:efficiency_combined}

\end{table*}

Since the similarity analysis has experimentally shown that \Approach~generated outputs align more closely with real repository documentation than existing methods for both tasks, we also aimed to evaluate its impact on efficiency by measuring response time and token usage.

\begin{figure}[ht]
  \centering
  \includegraphics[page=1,width=\linewidth]{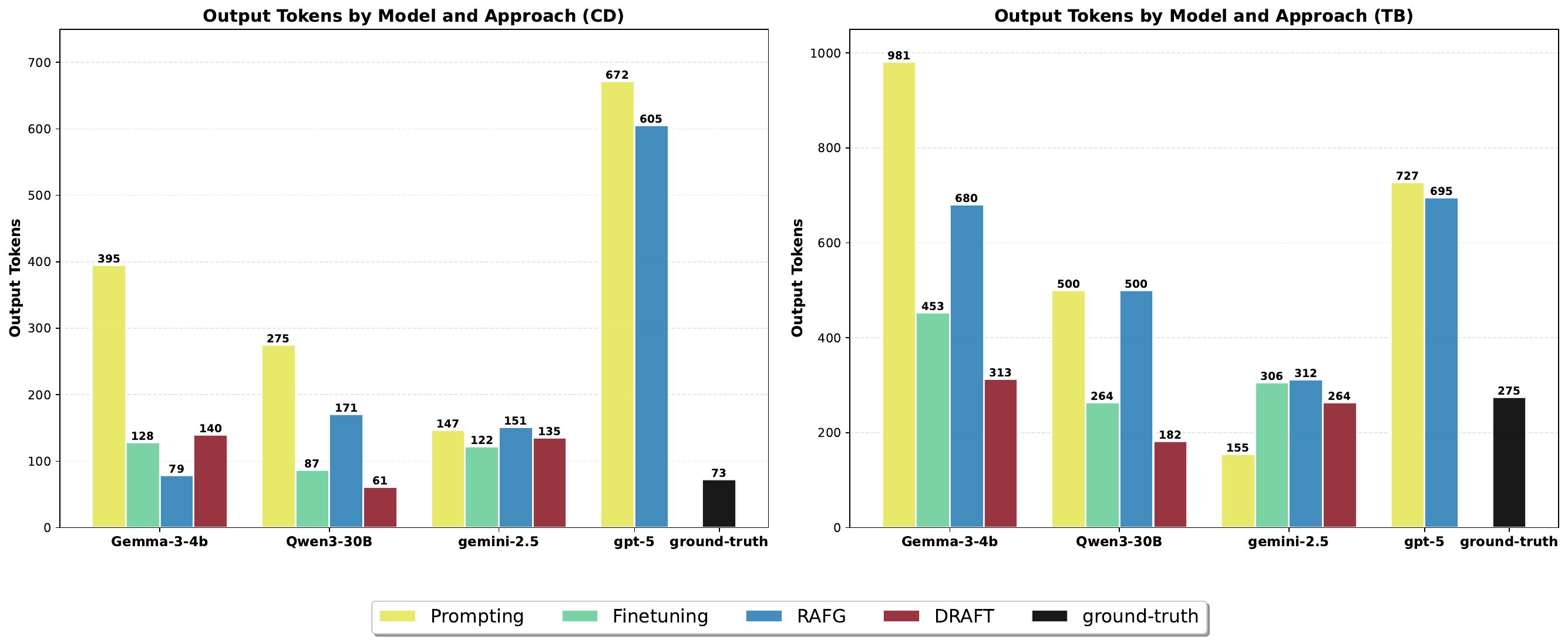}
  \caption{Mean output token counts per model and approach for Task 1 (left) and Task 2 (right), compared against ground truth lengths.}
  \label{fig:Output_Tokens}
\end{figure}

\begin{figure}[ht]
  \centering
  \includegraphics[page=1,width=\linewidth]{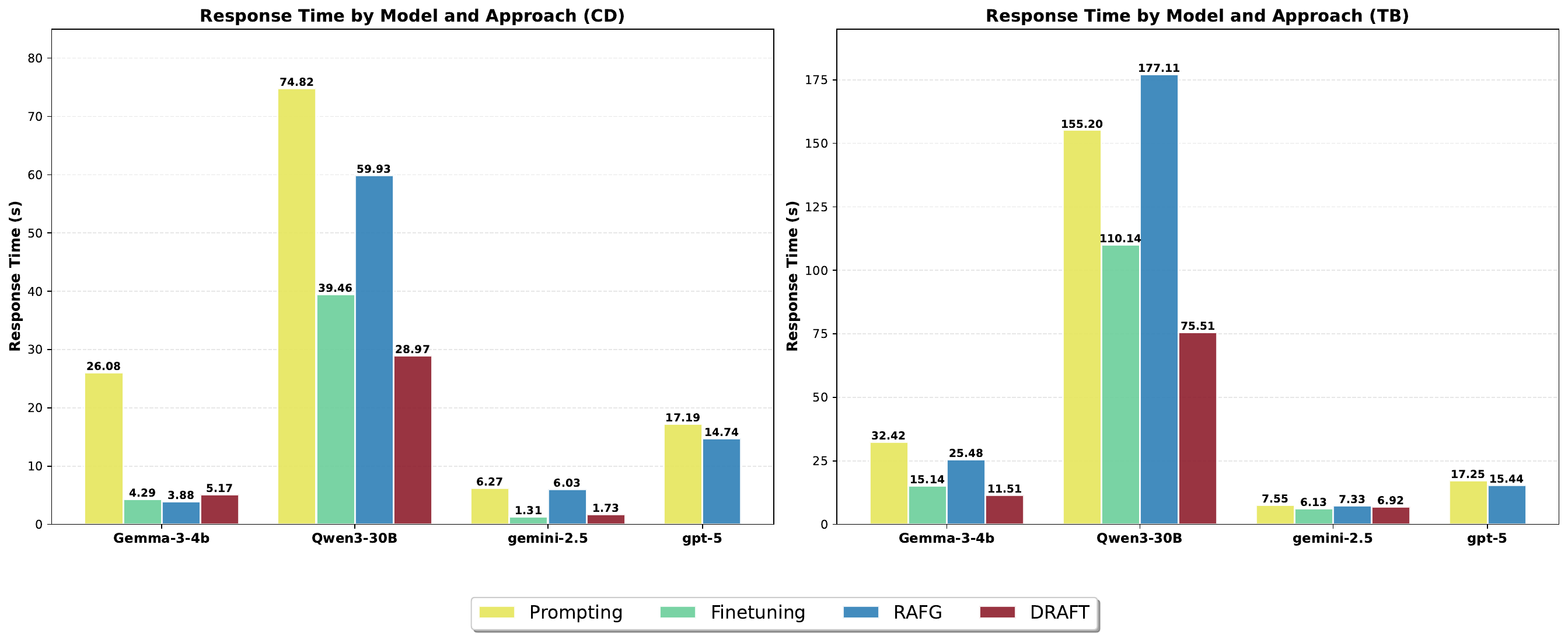}
  \caption{Mean response time (seconds) per model and approach for Task 1 (left) and Task 2 (right).}
  \label{fig:Response_Time}
\end{figure}

Token count serves as a proxy for cost, as most hosted models charge per token, while response time indicates system responsiveness, critical for real-time applications. We evaluated efficiency separately for both Task 1 (Context → Decision) and Task 2 (Title → Body) as they have different computational characteristics. During the experiments described in Section \ref{subsec:expCandidates}, we recorded token usage and response time for all model-approach combinations on the test set. We then extracted and analyzed this information to assess computational efficiency. The results are summarized in Table \ref{tab:efficiency_combined}, and visualized in Figures \ref{fig:Output_Tokens} and \ref{fig:Response_Time}.

\medskip
\noindent \textbf{Input Token Analysis}\\
As shown in Table \ref{tab:efficiency_combined}, \RAG~and \Approach~consume significantly more input tokens than prompting or \finetuning. This difference stems from their fundamental approach: \RAG~and \Approach~employ few-shot prompting, where multiple example input-output pairs are included in the prompt to guide the model, whereas prompting and \finetuning~use zero-shot prompts containing only a simple instruction and the input itself.

For Task 1, prompting uses 122 input tokens on average, while \RAG~and \Approach~use 527-559 input tokens (4-5× more) due to including 2 retrieved example C-D pairs. The contrast is even more dramatic for Task 2, where prompting uses only 8 input tokens on average since titles are very short, whereas \RAG~and \Approach~use 759-811 input tokens (100× more). This dramatic increase reflects both the brevity of the input title and the inclusion of 2 retrieved T-B pairs, which are substantially longer as they contain full ADR body examples.

\medskip
\noindent \textbf{Output Token Analysis}\\
As shown in Figure \ref{fig:Output_Tokens}, output token usage varies significantly by approach and task. For Task 1, ground truth decisions average 73 tokens. Prompting generates the longest outputs (147-672 tokens), often exceeding ground truth, while \finetuning~produces outputs closer to ground truth (87-128 tokens). \Approach~generates outputs similar to fine-tuning (61-140 tokens), with Qwen3-30B producing the most concise outputs at 61 tokens.

For Task 2, ground truth bodies average 275 tokens. Prompting again generates the longest outputs (155-981 tokens), with Gemma-3-4b producing particularly verbose bodies. \Finetuning~produces more controlled outputs (264-453 tokens), while \Approach~generates the most efficient outputs (182-313 tokens). This suggests \Approach~learns to generate appropriately detailed but concise ADR bodies that closely match the length distribution of real repository documentation.

\medskip
\noindent \textbf{Response Time Analysis}\\
As shown in Table \ref{tab:efficiency_combined} and Figure \ref{fig:Response_Time}, despite the additional overhead from prompt construction and longer inputs, \Approach~maintains competitive efficiency.

For Task 1, Qwen3-30B with \Approach~achieved a response time of 28.97 seconds, faster than its \finetuned~counterpart (39.46 seconds) despite using significantly more input tokens. Gemini-2.5 with \Approach~responded in 1.73 seconds, comparable to \finetuning~(1.31 seconds). GPT-5 with \RAG~showed 14.74 seconds, faster than prompting (17.19 seconds).

For Task 2, Qwen3-30B with \Approach~achieved a response time of 75.51 seconds, faster than both \finetuning~(110.14 seconds) and \RAG~(177.11 seconds). Gemma-3-4b with \Approach~responded in 11.51 seconds, significantly faster than both prompting (32.42 seconds) and \RAG~(25.48 seconds), while Gemini-2.5 with \Approach~maintained efficient performance at 6.92 seconds.

\medskip
\noindent \textbf{Cross-Task Efficiency Comparison}\\
Efficiency comparisons across tasks reveal clear differences in token usage and response time. Task 2 requires substantially more input tokens than Task 1 for \RAG~and \Approach~(759–811 vs. 527–559 tokens) because retrieved T–B pairs include full ADR bodies, whereas prompting and fine-tuning show minimal differences due to their simple zero-shot inputs.

For output tokens, Task 2 consistently generates longer outputs across all approaches, reflecting the more comprehensive nature of ADR bodies. Ground truth outputs average 275 tokens for Task 2 versus only 73 tokens for Task 1, a 3.8× difference. This pattern is maintained across all model–approach combinations: prompting generates 147–395 tokens for Task 1 versus 155–981 for Task 2, \finetuning~produces 87–128 versus 264–453 tokens, and \Approach~generates 61–140 versus 182–313 tokens.

Response time strongly correlates with output token count. Task 2 consistently incurs higher response times possibly due to longer generations. For example, on average Qwen3-30B with \Approach~requires 28.97 seconds to generate 61 tokens for Task 1 but 75.51 seconds to generate 182 tokens for Task 2, while Gemini-2.5 shows a similar increase (1.73 vs. 6.92 seconds). This pattern confirms output length as the primary driver of computational cost. Overall, \Approach~maintains competitive efficiency while producing outputs that more closely align with real repository documentation. Please note that the offline phases of all approaches have been excluded from this evaluation, as they represent one-time costs.

\section{Discussion}\label{sec:Discussion}

\subsection{Lessons Learned}\label{subsec:LessonsLearned}

\textbf{Impact of \Finetuning~and \RAG :}
Results of RQ1 show that while \RAG~improves performance by generating more context-aware outputs, its effectiveness varies across models and tasks. While for Task 1, \RAG~shows consistent improvements, for Task 2 (Title → Body), the benefits are more varied, with some models like Gemini-2.5 showing slight decreases in certain metrics.
We also observe that \finetuned~models generally outperform models relying solely on prompting or \RAG~across both tasks. This confirms that task-specific optimization helps LLMs generate better architectural artifacts. The improvements are particularly pronounced for Task 2, where the complexity of generating complete ADRs benefits more from domain-specific training.

\textbf{Model Capacity and \Approach~Complexity:} 
A key observation from our results is that \Approach -ing appears to work better with larger models. Performance gains seem to be highest and consistent for Gemini, our biggest \Approach -ed model. Whereas performance gains of Qwen are not that high, and is inconsistent accross metrics. For Gemma, our smallest model, performance gains by \Approach -ing is inconsistent and often dipping. This suggests that \Approach -ing, which combines the complexity of \RAFewshotG~with \finetuning, benefits from models with more parameters. Larger models appear better equipped to leverage the retrieved examples effectively while maintaining the domain-specific knowledge acquired through \finetuning.

\textbf{Response time depends on LLM:} 
Our analysis reveals that efficiency, particularly response time, is heavily dictated by the specific LLM used rather than the approach. Qwen consistently required a substantial amount of time, often an order of magnitude higher compared to the other models. In contrast, Gemini demonstrated the lowest response times, with Gemma and GPT falling in between. This variance might be attributable to the hardware-to-model combination, as the computational load of hosting the 30B parameter Qwen model on local hardware is significantly higher than that of the 4B parameter Gemma model or the optimized, API-based inference of Gemini.

\textbf{Correlation of Time and Output Tokens:}
We observed a positive correlation between response times and output token counts as can be inferred from Figure \ref{fig:Output_Tokens} and \ref{fig:Response_Time}. Task 2 consistently incurred higher response times than Task 1 primarily because it required generating significantly more tokens (averaging 182–313 tokens vs 61–140 tokens). This pattern holds true across different approaches as well. Whenever a specific approach resulted in a higher volume of generated tokens, such as the verbosity often observed with prompting, the response time also increased. This suggests that output length is a primary driver of response time.

\textbf{Aligning Model Output with Human-Like Brevity:} 
An important finding of our study is the relationship between output length and generation approach. Human written ADRs in our dataset are relatively concise, with a median Decision length of 73 tokens, and ADR Body length of 275 tokens. In contrast, prompting produces substantially more verbose outputs. When LLMs are \finetuned~or \Approach -ed on the dataset, they learn to generate more concise responses as reflected in Figure \ref{fig:Output_Tokens}. This behavior not only improves efficiency, but also better model outputs with dataset characteristics. Such alignment may also have contribute to improved evaluation scores.

\begin{rebuttalblue}
\textbf{Similarity to Real ADRs vs.\ Perceived Quality:} While \Approach~aligns best with real ADRs on automated metrics, human raters did not consistently prefer it: the authors favoured \Approach, the senior architect \RAG, and the junior architects saw little difference. A plausible explanation is that experienced practitioners value explicit rationale and alternatives, favouring the more elaborate \RAG~outputs. Whereas \Approach~mirrors the concise style of real ADRs, which reference based metrics reward but which often omits rationale, the most frequent criticism of \Approach~by the junior architects. Thus, closeness to real ADRs and usefulness to an architect are related but distinct goals.
\end{rebuttalblue}

\subsection{Implications for researchers}

\textbf{Generalizability Across Tasks and Domains:}
\Approach~effectively improves the quality of generated architectural artifacts across two distinct tasks: generating \Decision~and complete ADRs. A key direction for future research is testing the generalizability of \Approach~across other software architecture documentation tasks, such as generating architectural views, and broader software engineering domains.

\textbf{Task Complexity and Adaptive Configurations:}
The differential performance of \Approach~across both the tasks suggests that task complexity influences the benefits of combining retrieval and \finetuning. Future work should systematically investigate how task characteristics, such as input/output length, structural complexity, and information density affect \Approach 's performance. This could lead to adaptive \Approach~configurations optimized for different types of architectural documentation tasks.

\textbf{Advanced Retrieval and Fine-tuning Mechanisms:}
As our study reveals that retrieval and \finetuning~strategies increase LLMs' capability to generate architectural artifacts. Future research should explore more sophisticated retrieval mechanisms like knowledge graphs, hybrid retrieval combining dense and sparse methods, re-ranking strategies. \Finetuning~techniques, such as instruction tuning, parameter-efficient approaches beyond LoRA, and multi-task learning should also be explored. Additionally, investigating the optimal number of retrieved examples (k value) for different tasks and model sizes could further improve \Approach 's effectiveness.

\textbf{LLM Specific Efficiency Factors:}
Our efficiency analysis revealed that both response time and token usage are highly dependent on the specific LLM. Future research should investigate the factors that contribute to these variations, such as model architecture, inference optimization techniques, and hardware configurations.

\textbf{Output Length and Computational Cost:}
The strong correlation we observed between high response time and high output token counts (as shown in Figure \ref{fig:Output_Tokens} and \ref{fig:Response_Time},  and Tables \ref{tab:efficiency_combined} ) suggests that output length is a primary driver of computational cost. This relationship should be further investigated to develop strategies for predicting and optimizing inference costs. Research could investigate methods to better control the length of generated architectural artifacts while preserving output quality, as well as develop more efficient strategies for producing longer outputs.

\textbf{\Approach -ing Agents:}
Another research implication of this work lies in the emerging area of agentic AI systems, particularly in software engineering. Domain or task specific Agents are changing the world \cite{PICCIALLI2025128404}. Recent work on agentic AI systems has focused on agents that rely on RAG to access external knowledge during reasoning \cite{singh2025agenticretrievalaugmentedgenerationsurvey}. While RAG based agents have been widely explored, \finetuned~agents are only beginning to emerge. This raises an interesting research direction: exploring the impact of \Approach -ed Agents in specific Domains or tasks. \rebuttal{Particularly addressing engineering challenges such as context-window fragmentation, handling long retrieval traces, managing heterogeneous tasks within agentic workflows, efficiently storing and reusing \Approach-ed models, and selecting the most appropriate \Approach-ed agent for each task.}

\subsection{Implications for practice}

\textbf{Deployment options as per organizational Needs:}
Our study demonstrates that \finetuning~and \Approach -ing enhance the performance of LLMs in generating ADRs and helping in architectural decision making. This is important as the benefit of using ADRs as organizational practice is well established \cite{10.1007/978-3-031-70797-1_22}. On the other hand, organizations are increasingly seeking to optimize the cost of LLM based solutions, especially in terms of token consumption and infrastructure requirements. While large proprietary models often deliver strong performance, they can be prohibitively expensive and difficult to deploy in resource constrained environments.

Organizations can fine tune or \Approach~small LLMs and host them in house, benefiting from improved data privacy, personalization and cost efficiency while maintaining performance comparable to larger proprietary models. For those with budget or time constraints, \finetuning~alone may provide good balance of improved quality without the additional input token overhead of retrieval. On the other hand, for organizations using proprietary LLMs, \RAG~offers a middle ground improving performance over zero-shot prompting without requiring custom model training.

\textbf{Workflow Flexibility with Dual Task Support:}
The availability of two complementary tasks provides practitioners with flexibility in their workflows. Architects can use Task 1 (Context → Decisions) to explore decision alternatives during design, and Task 2 (ADR Title → Body) to transform finalized decisions into well structured ADRs, reducing the documentation burden that often discourages ADR adoption.


\begin{rebuttalblue}
\textbf{Task-Specific Performance Patterns:} Task~2 shows higher automated metric scores than Task~1 across all approaches, and \Approach{}'s gains over the baselines are larger for Task~2. Human raters, however, gave Task~2 the lowest Closeness ratings. The higher scores therefore likely reflect the shared structure and phrasing of ADR bodies rather than better reasoning. Hence Task~2 outputs are best used as structured drafts that an architect completes.
\end{rebuttalblue}

\textbf{Strategic Model Selection:}
Practitioners must balance model size, efficiency, and cost. Larger models such as Gemini and Qwen can deliver higher quality with \Approach . But efficiency varies widely across LLMs and hardware. Although \Approach~consumes substantially more input tokens than zero-shot prompting (4–5× for Task 1 and ~100× for Task 2) it produces more concise outputs aligned to human authored ADRs. Since response time is highly model dependent, organizations should benchmark candidate models on their own infrastructure to identify the best trade-off between quality, efficiency and cost.

\begin{rebuttalblue}
\textbf{Matching the Approach to the Purpose:} \Approach~is well suited for producing documentation that matches an organization's existing ADR style at low output cost, whereas for decision exploration, where rationale matters, \RAG~(rated most architecturally correct by the external experts) may be preferable. In both cases, outputs should be reviewed by an architect.
\end{rebuttalblue}

\textbf{Integration in Agentic AKM:}
As agentic systems are being proposed for AKM \cite{dhar2026agenticakmenrouteagentic}, a practical application of \Approach~is its integration into these emerging systems. Software agents can monitor development activities such as issue discussions, pull requests, and design documents to detect potential architectural decision points. When relevant contexts are identified, \Approach~can generate candidate decisions and assist in producing ADR documentation for review by architects. Approved ADRs can then be stored in organizational knowledge repositories and used to update the retrieval database, creating a continuous feedback loop that reduces documentation effort while improving the capture and reuse of architectural knowledge.

Overall our research indicates that \Approach , or it's sub approaches \RAG~and \Finetuning~can be used by architects as an assistant or co-pilot in drafting ADRs, following Russo et al. \cite{Copenhagen_Manifesto}.

\section{Threats to validity}\label{sec:threats_to_validity}

In this section, we discuss threats to the validity of our study, following the categorization provided by Wohlin et al. \cite{experimentSE}

\subsection{Internal Validity}
\textbf{Data Quality and Preprocessing:} As ADRs come in various formats, potential errors in data cleaning and standardization may influence model performance.
To mitigate this, we applied multi-stage data processing pipeline that includes string matching, regex based extraction, filtering and de-duplication. The details are in Section \ref{subsec:Dataset}.

Moreover, there might have been poor quality  ADRs in our dataset. However, the dataset is sourced from a prior MSR study \cite{adrs_data}, where the authors had conducted extensive screening, filtration, and eliminated low-quality ADRs. This involved a combination of automated and manual validation steps, ensuring overall dataset quality.

\textbf{Model Configuration: }The use of default values of LLM generation parameters, such as temperature, top-p, and top-k, represents another validity threat, as these parameters impact output quality. While alternative configurations were not tested, the chosen values align with best practices recommended by model providers. Future work could investigate the effects of tuning these parameters.

\textbf{Dataset Filtering:} The filtering of ADRs by size during dataset preparation may have excluded relevant instances, though the majority of ADRs were retained. To ensure quality and representativeness, we removed entries with body lengths outside 10–1000 tokens, required at least 90\% printable ASCII to filter non-English content, and excluded ADRs with over 20\% URL density. These thresholds were chosen based on token length analysis to ensure the dataset remained representative while being computationally manageable and focused on substantive architectural content rather than link collections or trivial entries.

\textbf{Evaluation Metrics Selection:} Another potential threat arises from the selection of evaluation metrics, as assessing the quality of text generation remains a complex challenge. To address this, we employed widely accepted NLP metrics (ROUGE-1, BLEU, METEOR, and BERTScore) that assess text quality from multiple perspectives including lexical overlap, precision, semantic similarity, and contextual embeddings.


\textbf{Stylistic Variations:} Inconsistencies in ADR writing styles may threaten validity through biased evaluations. Organizational or individual stylistic variations (e.g., precise/formal vs. descriptive) could result in lower scores for LLM generated outputs that conceptually align with manual ones but differ in expression. This risk was partially mitigated by incorporating diverse evaluation metrics that assess text quality from complementary angles, reducing the impact of any single metric's bias.

\begin{rebuttalblue}
\textbf{Human Evaluation:} Two evaluators were authors, which may bias ratings towards \Approach . We mitigated this through blinding, random shuffling, and an additional expert study, and report both groups separately. The expert study is small (14 instances, 20\% margin of error, three experts), and the two junior architects were merged into one rater, limiting statistical power and generalizability. Finally, ratings were subjective, as reflected by the fair to moderate inter-rater agreement, and exposure to the ground truth may have anchored correctness judgements.
\end{rebuttalblue}

\subsection{External Validity}

\textbf{Dataset Representativeness:} The dataset used in this study may not fully capture the diversity of architectural decision-making scenarios across different industries, such as embedded systems and cloud-native applications. To mitigate this threat, we used a dataset derived from an established MSR study \cite{adrs_data}, which followed rigorous procedures to crawl and compile ADRs from open-source projects on GitHub, ensuring representation of ADRs in general.

\textbf{Model Selection:} Another validity concern is the selection of LLMs, as only a subset of available LLMs could be evaluated. We mitigated this by selecting models based on the LM Arena leaderboard \cite{chiang2024chatbotarenaopenplatform}, including proprietary models and open-source models of different sizes. As detailed in Section \ref{subsec:LLMSelection}, this ensures coverage across architectures, parameter scales, and licensing, providing a representative snapshot of state of the art systems across cloud based and self hosted settings.

Similarly, retrieval quality may vary with different embedding models. To mitigate this, we adopted a vendor alignment strategy. Proprietary LLMs were paired with embedding models form the same vendor, while open source models used Qwen3-Embedding-8B, which ranked highest on the MTEB Retrieval leaderboard at the time of selection.

\subsection{Construct validity}

\textbf{Contextual Completeness:} In our study, the Context section of ADRs served as the input for generating Design Decisions in Task 1. While the Context section in some ADRs included pros/cons, consequences, or discussions of alternatives, this was not universally true across all ADRs in our dataset. This poses a threat to construct validity as the LLM might not receive comprehensive context to generate fully informed decisions. Moreover, architectural decisions are often shaped by interdependencies with other decisions or influenced by latent organizational knowledge not captured in individual ADR contexts. Our data processing pipeline includes a fallback mechanism that treats all content preceding the Decision section as context when explicit Context headings are absent, helping to capture more complete contextual information. Although \Approach~addresses this concern partially through its retrieval component by providing similar examples, the limitation remains. Therefore, generated \Decision~should serve as recommendations rather than final outputs, requiring human architects to incorporate broader context and judgment.

Similarly, ADR Titles serving as the input for generating complete ADRs in Task 2 poses a construct validity threat. While titles summarize core architectural intent, they often omit implicit organizational knowledge, architectural patterns, trade-offs, and project specific conventions. Architectural decisions are frequently influenced by historical dependencies and tacit practices that are not reflected in standalone titles. Although \Approach~partially mitigates this limitation through retrieval of similar ADRs, organization-specific nuances may still be missing.


\textbf{Efficiency Measurement Scope:} The efficiency analysis in Section \ref{subsec:rq3} focused exclusively on the online phase of inference, measuring response time and token usage during generation. The offline costs associated with \finetuning, \Approach -ing, and building vector databases were excluded as they represent one time costs. While this approach aligns with practical deployment scenarios where offline costs are amortized over many inference operations, it may limit the generalizability of results to scenarios where full lifecycle efficiency is considered. Future work could provide a more comprehensive cost analysis including offline phases.


\section{Related work}\label{sec:related_work}

With the rise of general-purpose AI assistants like ChatGPT and Gemini, the use of LLMs in Software Engineering (SE) has increased significantly \cite{10803393, nguyenduc2023generativeartificialintelligencesoftware}. LLMs are used for various SE tasks including Code Documentation \cite{bhattacharya2023exploringlargelanguagemodels}, Requirements Engineering \cite{cheng2025generativeairequirementsengineering}, creating database queries \cite{10803393}. Gao et al. \cite{10.1145/3712005} lists SE tasks that can be supported with Generative AI.
The software architecture community also recognizes the growing importance of AI in AKM \cite{Rivera_Gen_2024}.

\medskip
\noindent \textbf{LLMs for Software Architecture}

A vision paper by Eisenreich et al. \cite{10.1145/3643660.3643942} presents a direction for developing software architecture candidates semi automatically based on requirements using AI techniques. More recently Esposito et al. \cite{esposito2025generativeaisoftwarearchitecture} conducted a multivocal literature review synthesizing the use of Generative AI in software architecture across both peer-reviewed and grey literature. Their findings reveal that LLMs are primarily applied for architectural decision support (38\% of studies) and reverse engineering, with RAG adopted in 20\% of studies and fine-tuning in 12\%—confirming that the combination of these techniques remains underexplored. The review identifies architectural decision support as the dominant use case, directly motivating the contribution of \Approach. A complementary systematic literature review by Schmid et al. \cite{schmid2025softwarearchitecturemeetsllms} examined 18 primary studies on LLMs in software architecture and identified four application areas: reference architectures, classification and detection, extraction and generation, and assistant systems. Critically, they found that most approaches rely on simple prompting with decoder-only models, and that advanced techniques such as chain-of-thought prompting and RAG remain rarely explored, thus underscoring the novelty of our approach.

Jahić et al. \cite{10628428} provided a state of the practice analysis through an interview study and hands on LLM assisted architecture design, highlighting that LLMs are inconsistent when generating visual architectural solutions and that it often mixes abstraction levels without identifying what is important in design decisions. Bucaioni et al. \cite{bucaioni2025artificialintelligencesoftwarearchitecture} performed a systematic literature review on the integration of AI with Software Architecture, identifying automated design, trade-off analysis, and continuous documentation as active areas while calling for further research in context aware reasoning and real time architectural adaptation.

Soliman et al. \cite{10.1007/978-3-032-02138-0_7} evaluated seven LLMs across software architecture knowledge tasks and found that while LLMs encode useful architectural knowledge, their accuracy and response similarity vary significantly across models. A closely related effort by Maranhão et al. \cite{10.1145/3698322.3698324} proposed a Prompt Pattern Sequence approach to assist architects during architectural decision making. They defined structured patterns that guide the LLM through decision exploration. Sanchez et al. \cite{10.2139/ssrn.5775315} introduced a semi-automatic framework using an ensemble of LLMs to generate, de-duplicate, and cluster architectural design decisions through a consensus based mechanism. While these approaches helps in generating \Decision s, they do not incorporate retrieval from an existing ADR knowledge base or \finetuning~on domain specific data.

\medskip
\noindent \textbf{Assistants for Architectural Decision Making}

The domain of documenting ADDs and maintaining ADRs has always been challenging. Several studies have aimed to improve the quality of ADRs by examining large-scale usage and adoption patterns \cite{10.1007/978-3-642-39031-9_17}. However, a major challenge remains—the significant manual effort required to create and maintain ADRs \cite{10.1007/978-3-642-39031-9_17}. To address this issue, researchers have proposed various approaches. Díaz-Pace et al. \cite{10.1007/978-3-031-70797-1_21} developed \textit{ArchMind}, an AI assistant that uses a LLMs to guide, architects in formulating decisions. Marinho \cite{marinho2021archifyrecommenderarchitecturaldesign} presented \textit{Archify} for recommending ADRs based on software project requirements. However these approaches don't use any retrieval mechanism or any domain specific fine-tuning. Despite these efforts, widespread adoption of ADD documentation still remains a challenge.

\medskip
\noindent \textbf{RAG and Fine-tuning Combinations}

RAG has been changing the NLP landscape and different adaptions of RAG has been coming up as summarizer by Gao et al. \cite{gao2024retrievalaugmentedgenerationlargelanguage}. Zhang et al. \cite{zhang2024raftadaptinglanguagemodel} proposed \textit{RAFT} (Retrieval Augmented FineTuning), a recipe for domain-specific \Finetuned~RAG. It improved upon baseline RAG by distinguishing between relevant and irrelevant documents in the retrieved set. However they didn't incorporate few-shot prompting in their approach. Yu et al. \cite{yu2025visrag} integrated vision language models with RAG, introducing multimodal RAG for document understanding tasks

\RAFewshotG~has also been tried out by various research groups. Izacard et al. \cite{izacard2022atlasfewshotlearningretrieval} introduced Atlas, a retrieval-augmented language model designed for knowledge-intensive tasks with few-shot learning. Application of \RAFewshotG~is not limited to just NLP tasks. Zhao et al. \cite{zhao2024retrievalaugmentedfewshotmedicalimage} enhanced medical image segmentation by retrieving and leveraging similar annotated images from small datasets. However, to the best of our knowledge, none of the approaches perform domain specific \Finetuning~like \Approach.

\rebuttal{More complex retrieval architectures, such as Knowledge Graphs \cite{10903241}, re-ranking \cite{glass2022re2gretrievererankgenerate}, Graph RAG \cite{edge2025localglobalgraphrag}, and Self-reflection \cite{asai2023selfraglearningretrievegenerate} have also been shown to improve RAG performance. However exploring these within the \Approach~approach remains an avenue for future work.}

\medskip
\noindent \textbf{Prior Work on LLMs for ADR Generation}

Our prior work investigated whether LLMs could generate \Decision~from \Context~with respect to ADRs \cite{dhar2024llmsgeneratearchitecturaldesign}. We evaluated three approaches: zero-shot, few-shot, and fine-tuning with various LLMs. Our findings suggested that while LLMs cannot fully automate ADR generation, they can significantly reduce the effort required for ADD documentation.
\Finetuning~particularly enhanced the performance of smaller LLMs like Flan-T5, enabling them to produce results comparable to those of larger models. This suggested that smaller \Finetuned~models can be used in resource-constraint environments as they can be hosted with smaller organizations with minimum infrastructure. However advanced techniques like RAG was not explored in that work, and the study was limited to a single task (Context → Decision). The present work extends this by introducing a novel Task 2 (Title → Body) for complete ADR documentation, incorporating few-shot prompting, RAG, and \finetuning~within the \Approach~approach.

\section{Conclusion and Future work}\label{sec:conclusion}




ADRs are an important mechanism for capturing architectural knowledge, yet their adoption remains limited due to the effort required to document decisions. In this work, we investigated how LLMs can assist architects in generating ADR artifacts across two complementary tasks: generating Design Decisions from Decision Contexts (Context → Decision) and generating entire ADR from it's Titles (Title → Body).

To support these tasks, we examined three commonly used LLM based approaches, Prompting, \RAG, and \Finetuning . While each approach demonstrated certain strengths, our experiments revealed inherent limitations. Based on the insights, we proposed Domain-specific Retrieval Augmented Few-shot Tuning (\Approach ), a hybrid approach that integrates \RAG~with \Finetuning. 


\begin{rebuttalblue}
We evaluated \Approach~on 4,344 ADRs from open-source repositories using automated similarity metrics and a human evaluation with the authors and external architects. Across both tasks and multiple LLMs, \Approach~consistently generates outputs that align more closely with real-world ADRs while maintaining practical efficiency. Human raters, however, did not uniformly prefer it, with experienced practitioners valuing the richer rationale of \RAG. Overall, \Approach~provides an effective foundation for AI-assisted AKM, particularly for organizations seeking to reduce inference costs and dependence on proprietary models.
\end{rebuttalblue}

Several directions remain for future research. Firstly, this work focuses on generating ADR artifacts from structured inputs such as decision contexts and titles. Future work could extend the approach to derive ADRs from other software artifacts, including source code, commit histories, and issue discussions, enabling architectural knowledge to be captured more automatically during development.


\begin{rebuttalblue}
Second, future studies should extend our human evaluation to larger groups of practitioners and industry case studies, and explore rationale-aware generation to combine the alignment of \Approach~with the explanatory depth valued by experienced architects.
\end{rebuttalblue}

Moreover, improving the retrieval component of \Approach~through techniques such as re-ranking, hybrid retrieval, or architectural knowledge graphs may further enhance generation quality.

Finally, integrating \Approach~into interactive architectural tools or IDE based assistants could enable architects to iteratively draft and refine ADRs. Such systems may evolve toward agent based architectural knowledge management, where \Approach -ed agents assist in generating, validating, and maintaining ADRs as part of an Agentic AKM ecosystem.

\begin{rebuttalblue}

\section*{AI Use Disclosure}

In compliance with the \textbf{ACM Policy on Authorship} (\url{https://www.acm.org/publications/policies/new-acm-policy-on-authorship}), the authors disclose the use of generative AI tools (including ChatGPT) to assist with code generation and completion during the software development phase of this research. Every piece of AI-generated or modified code was rigorously reviewed, and verified by the authors before integration into the experimental replication package.


\end{rebuttalblue}

\bibliographystyle{unsrt}
\bibliography{references}





\end{document}